%% file: example_paper.tex
\documentclass{article}

\usepackage{microtype}
\usepackage{graphicx}
\usepackage{subfigure}
\usepackage{booktabs} 
\usepackage{caption}

\usepackage{hyperref}

\usepackage[accepted]{icml2025}

\usepackage{amsmath}
\usepackage{amssymb}
\usepackage{mathtools}
\usepackage{amsthm}

\usepackage[capitalize,noabbrev]{cleveref}

\theoremstyle{plain}

\theoremstyle{definition}

\theoremstyle{remark}

\usepackage[textsize=tiny]{todonotes}

\usepackage{tabularx}
\usepackage{multirow}
\usepackage[normalem]{ulem}
\useunder{\uline}{\ul}{}
\usepackage{tcolorbox}
\tcbuselibrary{breakable}
\usepackage{amssymb}

\usepackage{ifsym}

\icmltitlerunning{Perspective Paper on Principled Agent Engineering}

\begin{document}

\twocolumn[
\icmltitle{From Craft to Constitution: A Governance-First Paradigm \\for Principled Agent Engineering}



\icmlsetsymbol{equal}{*}

\begin{icmlauthorlist}
\icmlauthor{Qiang Xu}{yyy}
\icmlauthor{Xiangyu Wen}{yyy}
\icmlauthor{Changran Xu}{yyy}
\icmlauthor{Zeju Li}{yyy}
\icmlauthor{Jianyuan Zhong}{yyy}
\end{icmlauthorlist}

\icmlaffiliation{yyy}{CURE Lab., Dept. of CSE, The Chinese University of Hong Kong, Hong Kong S.A.R.}

\icmlcorrespondingauthor{Qiang Xu}{qxu@cse.cuhk.edu.hk}

\icmlkeywords{Agent Development; AriberOS; Governance-First Paradigm}


{\begin{center}
    \captionsetup{type=figure}
    \includegraphics[width=0.65\textwidth]{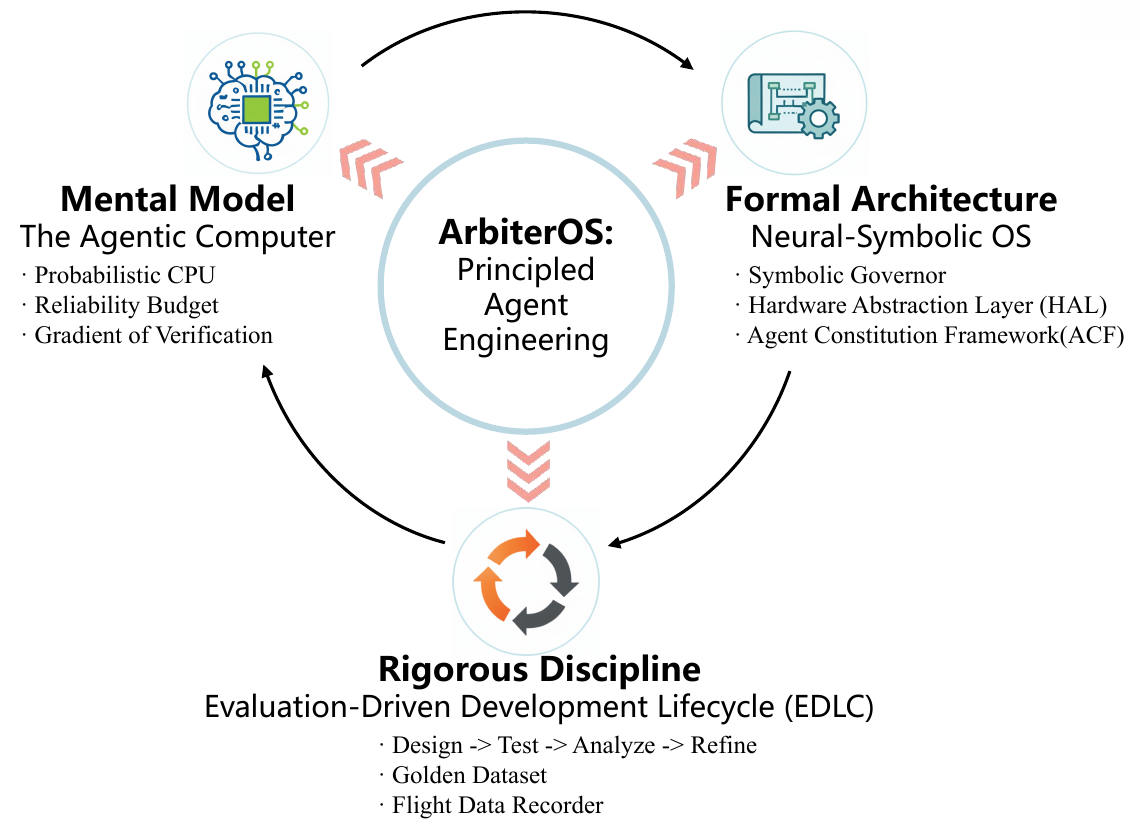}
    \vspace{5pt}
    \parbox{0.9\textwidth}{
    \small 
    \textbf{The ArbiterOS Paradigm.} An integrated framework for reliable AI agent engineering, combining a new \textbf{Mental Model} (the Agentic Computer) to understand probabilistic hardware, a \textbf{Formal Architecture} (a neuro-symbolic OS) to enforce safety, and a \textbf{Rigorous Discipline} (the Evaluation-Driven Development Lifecycle) for continuous verification. This constitution transforms agent development from a brittle craft into a principled engineering discipline.
}
    \label{fig:framework}
\end{center}}

]



\printAffiliationsAndNotice{}  

\begin{abstract}


The advent of powerful Large Language Models (LLMs) has ushered in an ``Age of the Agent,'' enabling autonomous systems to tackle complex goals. However, the transition from prototype to production is hindered by a pervasive ``crisis of craft,'' resulting in agents that are brittle, unpredictable, and ultimately untrustworthy in mission-critical applications. This paper argues this crisis stems from a fundamental paradigm mismatch---attempting to command inherently probabilistic processors with the deterministic mental models of traditional software engineering. To solve this crisis, we introduce a \textbf{governance-first paradigm} for principled agent engineering, embodied in a formal architecture we call \textbf{ArbiterOS}.


This paradigm provides control by introducing the \textbf{Agentic Computer}, a mental model that reframes the LLM as a ``Probabilistic CPU.'' A neuro-symbolic \textbf{Hardware Abstraction Layer (HAL)} that decouples durable agent logic from the underlying model manages this volatile hardware, which is governed by the formal \textbf{Agent Constitution Framework (ACF)} instruction set and continuously verified by the rigorous \textbf{Evaluation-Driven Development Lifecycle (EDLC)}.


Ultimately, ArbiterOS provides a coherent blueprint to move beyond the craft of prompting and begin the discipline of principled agent engineering. Through its governance-first architecture, it provides the foundations for building agents that are not only reliable, auditable, and secure, but also \textbf{portable, compliant by design, and organizationally scalable}---transforming fragile prototypes into robust, production-ready systems.

\end{abstract}

\input{content/Introduction}
\input{content/relted_work}
\input{content/Sec2}

\input{content/Sec3}
\input{content/Sec4}
\input{content/Sec5}
\input{content/Sec6}

\input{content/Sec7}

\section*{Conclusion and Future Work}
\label{sec:conclusion}

This paper has argued that the path forward for agentic AI is not a more clever prompt, but a paradigm shift: from a ``crisis of craft'' to a true engineering discipline. We have laid out a blueprint for that discipline: the \textbf{ArbiterOS paradigm}. It introduces a `Kernel-as-Governor' model as a necessary complement to orchestration-focused AIOS frameworks, built upon the formal mental model of the \textbf{Agentic Computer}, a neuro-symbolic \textbf{Hardware Abstraction Layer (HAL)}, and the rigorous \textbf{Evaluation-Driven Development Lifecycle (EDLC)}. The purpose of this unified framework is to provide the architectural and methodological foundation to engineer agents that are not merely capable, but provably reliable, auditable, and secure. Moreover, the paradigm's formal structure is the key that unlocks the next level of maturity for the field, enabling portability, organizational scalability, and compliance by design.

The journey from craft to an engineering discipline requires a collective shift in mindset. This work has sought to enable that shift by providing a model to ask the right questions, an architecture to build the answers, and a discipline to verify them. This paper is therefore not just a proposal, but a call to action to begin the real work of engineering constitutions for our agents. As a perspective paper, our goal has been to provide this architectural blueprint; the immediate next step is to ground the paradigm in practice through a reference implementation that will enable the rigorous validation of the reliability gains and performance trade-offs discussed.

\subsection*{Future Work: From Bottlenecks to Governed Autonomy}

The adoption of a formal paradigm like ArbiterOS does not end the engineering challenges; rather, it illuminates the next set of critical research frontiers. By providing a stable architectural foundation, it allows the field to systematically address the bottlenecks that stand in the way of truly autonomous systems. We identify three primary horizons for future work.

\paragraph{Priority 1: Overcoming the Machine Bottleneck via Automated Optimization.} The ``Abstraction Tax'' of a governance layer, particularly its impact on latency, is a primary engineering challenge to be solved. The long-term vision is to build \textbf{`Agent Compilers.'} As introduced in Section~\ref{subsec:compilers}, these tools would treat the Agent Constitution as an Intermediate Representation (IR) to automatically rewrite execution graphs and pre-compile policies, with the primary goal of minimizing end-to-end latency. This research is critical to making governed agents practical in latency-sensitive applications.

\paragraph{Priority 2: Overcoming the Human Bottleneck via an Automated Evaluation Lifecycle.} 
While performance is a machine-level bottleneck, we believe the most urgent R\&D challenge for scalability is the \textbf{human bottleneck}. As highlighted in Section~\ref{sec:discipline}, this manifests as the \textbf{``Oracle Problem''}: the reliance on manual, expert-driven failure analysis and test case curation for the Golden Dataset. Overcoming this is the critical path to making principled agent engineering truly scalable. We therefore prioritize research into a self-improving evaluation ecosystem, including: \textbf{Automated Red-Teaming} to discover novel failure modes, and \textbf{Automated Test Case Generation} from production failure traces.

\paragraph{The Next Frontier: Extending the ACF for the Era of Experience.}
Once the foundational challenges of \textit{reliable execution} and \textit{scalable learning} are met, the next great frontier is to enable governed autonomy, moving toward the ``Era of Experience'' discussed in Section~\ref{sec:knowledge_vs_experience}. This requires extending the governance paradigm from the execution of pre-defined tasks to the very processes of autonomous learning, collaboration, and human interaction. Realizing this vision requires extending the Agent Constitution Framework with new, forward-looking cores that transform opaque, probabilistic learning processes into governable, auditable operations.

\begin{itemize}
    \item \textbf{The Adaptive Core: Governed Self-Improvement.} As agents acquire the ability to autonomously learn, alignment during adaptation becomes the central concern. We propose the \textbf{Adaptive Core}, an instruction set that formalizes learning and self-modification as governable operations. This core provides the "meta-methods" for how knowledge acquisition, skill refinement, and preference learning can proceed under auditable policy constraints, enabling safe, continual improvement.

    \item \textbf{The Social Core: Multi-Agent Coordination.} The transition from single-agent control to multi-agent ecosystems introduces new challenges of communication and emergent risk. We propose the \textbf{Social Core}, which offers a governable instruction set for inter-agent interaction. Its policies would ensure that communication protocols are enforced, negotiations remain bounded, and collective decision-making processes yield auditable outcomes.

    \item \textbf{The Affective Core: Human-Agent Alignment.} For agents to operate as effective teammates rather than mere tools, they must be capable of reasoning about human cognitive and emotional states. We propose the \textbf{Affective Core}, which formalizes socio-emotional reasoning within a governable framework. This would enable agents to interpret implicit user intent, adapt communicative behavior, and manage trust dynamics in alignment with human needs.
\end{itemize}

Solving these challenges across all three frontiers---performance, learning, and autonomy---will be critical to breaking the final bottlenecks. It will enable the transition from a systematic craft to a truly mature discipline capable of building the next generation of AI systems. Our goal is to provide the architectural foundation that allows the development of reliable, adaptive AI agents to finally become what all mature engineering disciplines aspire to be: productively boring. A more detailed exploration of this forward-looking vision is presented in Appendix~\ref{app:sutton}.

\bibliography{example_paper}
\bibliographystyle{icml2025}

\newpage
\appendix
\onecolumn

\input{content/Appendix}




\end{document}

%% file: content/Introduction.tex
\section{Introduction}
\label{sec:intro}


The advent of powerful Large Language Models (LLMs), catalyzed by foundational innovations like the Transformer architecture~\citep{vaswani2017attention}, has ushered in the `Age of the Agent'---a new paradigm in software where autonomous systems can reason, plan, and act to achieve complex goals~\citep{masterman2024landscapeemergingaiagent}. The potential is transformative, promising a future of personalized Socratic tutors, AI employees that autonomously manage supply chains, and truly capable digital scientists designing novel proteins---applications once the domain of science fiction~\citep{xi2025rise}. This vision, however, is colliding with the harsh engineering reality of a foundational challenge that plagues nearly every team attempting to move complex agents from curated demos to production environments: a pervasive `crisis of craft.'\footnote{This `crisis of craft' is not unique to agentic systems; it mirrors historical paradigm shifts in other fields, such as the `software crisis' of the 1960s, where the failure of ad-hoc programming at scale necessitated the formation of software engineering as a formal discipline.}



This crisis is not merely anecdotal; it is being systematically quantified. For instance, a recent large-scale benchmark evaluated leading LLM agents on consequential real-world tasks like trip planning and online shopping. The study found that even state-of-the-art models such as Gemini-2.5-Pro and GPT-4o, achieved a \textbf{success rate of only 30\%}, revealing a significant gap between demonstrated capabilities and production-ready reliability~\citep{TheAgentCompany2024}. This failure manifests as a consistent and painful set of symptoms: agents are \textbf{brittle}, shattering when faced with slight variations from their training data~\citep{shah2024agents}; they are \textbf{unpredictable}, with non-deterministic behaviors that turn debugging into a costly exercise in guesswork; and they are ultimately \textbf{unmaintainable}, rendering formal safety guarantees impossible and turning the critical task of demonstrating regulatory compliance into speculation~\citep{wang2025surveyresponsiblellmsinherent}.


These symptoms stem from a common root cause: the reliance on natural language as the primary programming interface. While powerful, this interface introduces a level of ambiguity and instability that stands in stark contrast to the precision of formal programming languages~\citep{sahoo2025systematicsurveypromptengineering}. This instability leads directly to unmaintainability, as agents devolve into a tangled web of elaborate prompts where a minor change can cause unforeseen, catastrophic failures---a phenomenon known as ``prompt drift"~\citep{chen2023chatgptsbehaviorchangingtime}. For instance, a seemingly innocuous tweak to an agent's prompt, intended to make its tone more professional, could inadvertently cause it to start misclassifying user support tickets, leading to a silent but critical failure in a customer service workflow. 

Furthermore, this same informal interface creates significant insecurity by exposing poorly understood attack surfaces vulnerable to techniques like prompt injection~\citep{chen-etal-2025-defense}. In short, we are attempting to build mission-critical infrastructure, akin to a nation's power grid, using the bespoke methods of a medieval artisan---an approach fundamentally mismatched to the scale, reliability, and safety the task demands.

This paper posits that this crisis stems from a deeper, fundamental paradigm mismatch: \emph{attempting to command an inherently probabilistic processor using the deterministic mental models of traditional software engineering}. The LLM is a quintessential ``Software 2.0" component~\citep{Andrej2027software}; its behavior is learned from data rather than being explicitly programmed, a paradigm shift with profound implications for traditional software engineering practices~\citep{Dig2021Understanding}. The artisanal approach of a ``prompt wizard" is a brittle attempt to force deterministic behavior from this probabilistic system, treating it like traditional code and failing to manage its inherent uncertainty.

This mismatch has forced practitioners to seek more robust solutions through industrialized approaches, evidenced by a wave of recent releases from major industry leaders. Chief among these are OpenAI's AgentKit~\citep{openai2025agentkit}, Microsoft's Agent Framework (MAF)~\citep{microsoft2025framework}, and Google's agent development stack under the Gemini Enterprise umbrella~\citep{google2025enterprise}. These platforms represent a critical maturation of the field's \textbf{execution mechanisms}, providing the essential tooling and orchestration for building complex agentic workflows.

While these advances are foundational, they do not resolve the root cause identified earlier. They provide the ``how'' of execution, but not the ``what'' of guaranteed, reliable behavior. What is missing is a complementary \textbf{governance paradigm}: a formal architectural approach for managing the probabilistic outputs of LLMs. A truly robust solution demands such an architecture---one that enforces reliability by design.

To fill this critical gap between powerful execution mechanisms and the need for reliable control, this paper introduces a \textbf{governance-first paradigm} for principled agent engineering, embodied in a formal Operating System (OS) architecture we call \textbf{ArbiterOS}. This paradigm is not a replacement for existing execution engines, but rather the governing layer that makes them safe and production-ready. By providing the systemic enforcement mechanisms that informal principles lack, this paradigm represents the next logical step in the field's evolution from craft to an engineering discipline, crystallizing an emerging industry-wide trend toward formalization. This integrated paradigm is built upon three pillars that form a new, coherent approach to agent development:

\vspace{-5pt}

\begin{itemize}
    \item \textbf{A New Mental Model to Think With:} We begin by reframing the agentic system as an \textbf{Agentic Computer}, with the LLM as its core ``Probabilistic CPU\footnote{While the LLM is software, from the perspective of an agent architect who cannot modify its weights, it functions as the system's core processor.}." This model allows us to reason about reliability as a systems management problem and apply time-tested principles from classical computer architecture.
    \item \textbf{A Formal Architecture to Build With:} We propose a neuro-symbolic architecture centered on a deterministic \textbf{Symbolic Governor}, which acts as the system's trusted OS kernel. The Governor's role is to provide fine-grained, intra-agent governance---imposing rules and safety checks directly on the agent's workflow. 

    This establishes our ``Kernel-as-Governor'' paradigm, a distinct approach focused on internal reliability. It contrasts with the broader AI Agent Operating System (AIOS) field, which has primarily adopted an orchestration-centric ``Kernel-as-Scheduler'' model for managing interactions \emph{between} multiple agents. To achieve this, our architecture provides two key components:
    \begin{itemize}
        \item A \textbf{Hardware Abstraction Layer (HAL)} that acts as a universal adapter for the volatile LLM. It decouples the agent's core logic from the specific, ever-changing details of the underlying model, ensuring the agent is portable and maintainable.
        \item The \textbf{Agent Constitution Framework (ACF)}, a formal instruction set architecture (ISA) for governance. The ACF provides the clear, machine-readable rulebook that the Governor uses to enforce reliability and safety policies.
    \end{itemize}
    \item \textbf{A Rigorous Discipline to Verify With:} We operationalize this paradigm through the \textbf{Evaluation-Driven Development Lifecycle (EDLC)}. This cyclical discipline is centered on improving agent performance against version-controlled benchmarks, transforming reliability from a vague aspiration into a measurable, data-driven engineering goal.
\end{itemize}


\textit{The ultimate goal of any mature engineering discipline is to make the construction of complex, powerful systems predictable, reliable, and, ultimately, \textbf{productively boring}.} While these three pillars---a new Mental Model, a Formal Architecture, and a Rigorous Discipline---draw on established principles, their synthesis into a single, coherent OS paradigm for probabilistic hardware is this paper's primary contribution. It is an integrated framework designed to transform the disparate concepts of agent development from a brittle craft into precisely that: a principled and predictable engineering discipline.
\newpage

The primary contributions of this work are therefore to:
\begin{itemize}
    \item Formalize the engineering challenges of agentic systems through the \textbf{Agentic Computer} model.
    \item Propose \textbf{ArbiterOS}, a neuro-symbolic operating system paradigm that provides a Hardware Abstraction Layer for auditable, policy-driven governance over this new class of hardware.
    \item Define the \textbf{EDLC}, a formal development discipline for systematically building and maintaining verifiably reliable agents in response to the challenges of ``non-stationary hardware."
\end{itemize}

Furthermore, this paper argues that the same architectural principles required to solve today's reliability crisis are also the essential foundation for the next frontier of AI: \textit{enabling autonomous agents to learn safely and continually from experience}. This governance-first paradigm provides not just a solution for the present, but a necessary blueprint for the future.

The remainder of this paper is structured as follows. Related works are discussed in Section~\ref{sec:related_work}. Section~\ref{sec:computer&budget} introduces the Agentic Computer model and its core engineering challenges. Section~\ref{sec:arbiteros} details the ArbiterOS paradigm and its neuro-symbolic architecture. Section~\ref{sec:acf} defines the Agent Constitution Framework. Section~\ref{sec:governancePrinciple} provides an illustrative walkthrough of the paradigm in action. Section~\ref{sec:discipline} describes the EDLC, and Section~\ref{sec:situatingArbiterOS} situates ArbiterOS within the broader agentic ecosystem. Finally, we conclude this perspective paper and discuss future work in Section~\ref{sec:conclusion}.

%% file: content/relted_work.tex
\section{Related Work and Motivation}
\label{sec:related_work}

\begin{table*}[!t]
\centering
\small
\caption{A comparative analysis of ArbiterOS against adjacent frameworks, highlighting its unique focus on intra-agent governance and architectural enforcement.}
\label{tab:framework_comparison}
\vspace{5pt}
\renewcommand{\arraystretch}{1.5}
\begin{tabular}{@{}lp{2.6cm}p{2.8cm}p{2.8cm}p{3.9cm}@{}}
\toprule
\textbf{System/Framework} & \textbf{Primary Focus} & \textbf{Enforcement Mechanism} & \textbf{Core Abstraction} & \textbf{Key Limitation Addressed by ArbiterOS} \\
\midrule

\textbf{\begin{tabular}[t]{@{}l@{}}LangChain/\\LangGraph\end{tabular}} & Low-Level Execution Mechanism & Imperative Code (Developer's responsibility) & Agent as a ``stateful graph" & Lacks explicit governance mechanism. Relies on the developer to correctly apply governance nodes. \\

\textbf{OpenAI AgentKit} & Developer Velocity \& Integrated Platform & Application-Level Logic (Visual nodes, e.g., ``Guardrails,'' ``Approval'') & Agent as a ``visually composed workflow'' & Enforcement is a configurable part of the workflow, not a separate, guaranteed architectural layer. Relies on the developer to correctly apply governance nodes. \\

\textbf{\begin{tabular}[t]{@{}l@{}}Microsoft Agent\\ Framework\end{tabular}} & Inter-Agent Orchestration \& Enterprise Middleware & Middleware \& SDK-level Logic (Filters, telemetry hooks, explicit graphs) & Agent as a ``component in a workflow'' (probabilistic or deterministic) & Lacks a formal governance ISA. Governance is implemented via code/middleware, not enforced by a distinct kernel that separates the governor from the governed. \\

\textbf{\begin{tabular}[t]{@{}l@{}}Google Gemini\\ Enterprise / ADK\end{tabular}} & Integrated Platform \& Open Ecosystem Tools & SDK-level Logic \& Middleware (``deterministic guardrails") & Agent as a ``component in a managed runtime" & Lacks a formal governance ISA and the architectural separation of governor and governed. Relies on developer-implemented controls within the SDK. \\

\textbf{AIOS, KAOS} & Inter-Agent Orchestration \& Resource Management & OS-level Scheduling & Agent as a scheduled ``process'' or ``app'' & Lacks fine-grained governance of the agent's internal, probabilistic execution; treats the agent as a black box. \\


\textbf{TrustAgent} & Intra-Agent Safety \& Trustworthiness & Application-Level Logic & Agent as a self-regulating entity consulting a ``constitution'' & Enforcement is not architecturally guaranteed; relies on the agent's own probabilistic logic to follow the rules, which can be subverted. \\


\textbf{CoALA} & Conceptual Framework for Agent Specification & N/A (Descriptive Framework) & Agent as a set of memory, action, and decision modules & Provides a language for \textit{describing} agents but does not offer an \textit{enforceable architecture} or engineering discipline for building them reliably. \\

\midrule[1.5pt] 

\textbf{ArbiterOS (This Work)} & \textbf{Intra-Agent Governance \& Reliability} & \textbf{Architectural (Deterministic OS Kernel)} & \textbf{Agent as a governed ``Agentic Computer''} & Provides the missing \textbf{architectural enforcement layer} for intra-agent reliability, making other paradigms (e.g., collaboration) safer to deploy. \\

\bottomrule
\end{tabular}
\label{table:compare_OS}

\end{table*}

\subsection{The Rise of LLM-Based Agents and the Reliability Gap}

The rise of powerful LLMs has catalyzed the development of autonomous AI agents capable of complex task execution through sophisticated pipelines of planning, reasoning, and tool-calling~\citep{masterman2024landscapeemergingaiagent}. 

However, as these agents move from research prototypes to real-world deployment, a critical gap has emerged between their impressive capabilities and their operational reliability. This ``reliability gap" is a central theme in the Software Engineering for AI/ML (SE4AI) community, which has identified significant, recurring challenges in building, operating, and maintaining AI-based systems. As noted in a survey by~\citep{10.1145/3487043}, the most studied properties of these systems are dependability and safety, yet data-related issues and a lack of mature engineering practices remain prevalent challenges. Recent security surveys further underscore this gap, highlighting that agents face significant threats from untrusted inputs and complex internal states, making trustworthiness a paramount concern~\citep{10.1145/3716628, shi2025towards}.

In response, various approaches have sought to bridge this reliability gap. Frameworks like TrustAgent, for instance, integrate safety `constitutions' directly into the agent's operational loop to guide its behavior~\citep{hua2024trustagent}. Meanwhile, other research explores the application of formal methods to verify specific properties of agents, advocating for a fusion of LLMs with mathematically rigorous techniques to produce more reliable outputs~\citep{zhang2024fusionlargelanguagemodels}. While these approaches represent important steps, they often address specific aspects of the problem in a fragmented manner. They provide valuable guidelines or point-solutions but do not yet offer a unified, architectural paradigm for the systematic engineering of reliable agents from the ground up. What is missing is a coherent framework that treats the LLM as a new class of hardware and provides an OS-level abstraction needed for robust, systemic governance. 

\subsection{The Emergence of Integrated Agentic Platforms}
\label{sec:emerging_platforms}


The ``crisis of craft'' we identify is no longer a prospective challenge but a present-day engineering reality. This fact is powerfully validated by the recent, large-scale introduction of comprehensive agent development frameworks from key industry leaders. Chief among these are OpenAI's AgentKit~\citep{openai2025agentkit}, Microsoft's Agent Framework~\citep{microsoft2025framework}, and Google's Gemini Enterprise stack~\citep{google2025enterprise}. The very existence of these platforms represents a massive, collective industry investment to solve the exact problems of brittleness and scalability born from ad-hoc, prompt-centric approaches. They seek to replace a landscape of ``fragmented tools'' with cohesive platforms that unify research innovation with ``enterprise-grade foundations.''


These frameworks, while sharing a common goal, reveal a strategic spectrum in the market, now defined by three primary approaches: the integrated platform, the open-source ecosystem, and the enterprise ecosystem.

\begin{itemize}
    \item \textbf{The Platform Paradigm (Archetype: OpenAI AgentKit):} AgentKit exemplifies the integrated platform approach. It offers a highly integrated, visually-driven ``platform-as-a-service'' that prioritizes developer velocity and a seamless user experience, but which is deeply coupled to the broader OpenAI ecosystem. The trade-off is potential ecosystem lock-in.

    \item \textbf{The Open-Source Ecosystem Paradigm (Archetype: Microsoft Agent Framework):} In contrast, Microsoft's Agent Framework embodies the open ecosystem approach. It is positioned as an extensible, interoperable SDK and runtime designed to serve as a foundational, middleware layer, promoting interoperability through open standards, e.g., A2A (Agent-to-Agent communication) and MCP (Model Context Protocol). The trade-off is greater inherent complexity.

    \item \textbf{The Enterprise Ecosystem Paradigm (Archetype: Google Gemini Enterprise):} Google's agent stack represents a hybrid approach. It combines a managed, enterprise-grade runtime (\textbf{Agent Engine}) and a central governance plane with an open-source, model-agnostic SDK (\textbf{Agent Development Kit}). This model aims to offer enterprise-level security and observability while still fostering interoperability within the broader ecosystem, for instance, through the A2A communication protocol.
\end{itemize}

The expanded comparative analysis in Table~\ref{table:compare_OS} provides a high-level map of this new, more nuanced ecosystem, situating these frameworks alongside their predecessors and highlighting the architectural gap ArbiterOS is designed to fill.

To properly deconstruct these complex systems and identify this architectural gap, we must first introduce a crucial distinction that cuts across all three paradigms: the difference between an execution \textit{mechanism} and a governance \textit{paradigm}.

\begin{table*}[t!]
\centering
\caption{A Comparative Analysis of Agent Governance Mechanisms}
\label{tab:governance_comparison}
\begin{tabularx}{\textwidth}{@{} l >{\raggedright\arraybackslash}X >{\raggedright\arraybackslash}X >{\raggedright\arraybackslash}X @{}}
\toprule
\textbf{Feature} & \textbf{OpenAI AgentKit} & \textbf{Microsoft Agent Framework / Google ADK} & \textbf{ArbiterOS (This Work)} \\
\midrule
\textbf{Paradigm} & 
Application-Level Governance & 
Middleware-Level Control & 
\textbf{Kernel-Level Governance} \\
\midrule
\textbf{Mechanism} & 
Visual ``Guardrail'' \& ``Approval'' nodes manually placed in the agent's workflow graph. & 
Imperative filters, telemetry hooks, and explicit graph logic (or SDK-level ``guardrails") defined in code by the developer. & 
Declarative policies (e.g., in YAML) automatically enforced by a trusted, formally separated runtime (the Kernel). \\
\midrule
\textbf{Guarantee} & 
\textit{Developer-Dependent.} Relies on the developer to correctly configure and connect governance nodes. & 
\textit{Developer-Dependent.} Relies on the developer to correctly implement governance logic within the SDK. & 
\textit{Architecturally Enforced.} Provides a guaranteed separation of the governor from the governed, making compliance an intrinsic property of the system. \\
\bottomrule
\end{tabularx}
\end{table*}

\subsection{An Architectural Lens: Execution Mechanism vs. Governance Paradigm}

A common point of confusion in the rapidly evolving agentic landscape is the conflation of two distinct architectural layers: the low-level \textbf{execution mechanism} and the high-level \textbf{governance paradigm}. Understanding this distinction is the key to identifying the root cause of the ``crisis of craft'' and appreciating the unique contribution of ArbiterOS.

\subsubsection{Execution Mechanisms: The `Bricks and Mortar'}

An execution mechanism is the underlying substrate that provides the stateful, cyclical, and graph-based control flow for running an agent. Contemporary frameworks like \textbf{LangGraph}~\cite{langchain2023langgraph} exemplify this layer perfectly. They provide the essential, low-level primitives---the ``bricks and mortar''---for constructing complex workflows as a graph of nodes (computational units) and edges (control paths). The powerful runtimes within Microsoft's Agent Framework and OpenAI's AgentKit are, at their core, sophisticated and robust implementations of such mechanisms.

However, these mechanisms are, by design, \textbf{agnostic}. They provide the engine for execution but are not opinionated about the safety, reliability, or auditability of the workflow. An execution engine will faithfully run an unsafe workflow just as readily as a safe one.

\subsubsection{Governance Paradigms: The `Blueprint and Building Codes'}

A governance paradigm, by contrast, is the formal architectural and enforcement framework that transforms an arbitrary workflow into a reliable, auditable system. It provides the ``blueprint and building codes'' that dictate how the mechanisms can be safely used. Its purpose is not merely to execute a graph, but to ensure the execution adheres to a set of formally specified, human-defined reliability and safety guarantees.

This distinction is best understood through a classic analogy from computer science: the CPU versus the Operating System.
\begin{itemize}
    \item A \textbf{CPU (The Mechanism)} provides a raw, powerful set of instructions. It is an unopinionated execution engine.
    \newpage
    \item An \textbf{Operating System (The Paradigm)} provides the essential governance layer. It manages processes, enforces permissions, handles faults, and provides the structured environment that makes the raw power of the CPU safe and useful for complex applications.
\end{itemize}


Frameworks like LangGraph are the agentic equivalent of the CPU's raw instruction set. Crucially, however, not all ``Operating Systems" serve the same function, which we detail in the following subsection.

\subsection{Paradigms in Agentic Systems Architecture}

Armed with this architectural lens of mechanism versus paradigm, we can now deconstruct the foundational patterns---such as the `Kernel-as-Scheduler' or `Application-Level Governance'---that are being implemented and extended within today's most advanced frameworks.

\subsubsection{The `Kernel-as-Scheduler': Inter-Agent Orchestration}

Recognizing the need for systemic management, the paradigm of the \textbf{AI Agent Operating System (AIOS)} has emerged, galvanized by early conceptual work establishing the core analogy of ``LLM as OS, Agents as Apps''~\citep{ge2023llm}. The first wave of these systems adopted a \textbf{`Kernel-as-Scheduler'} model, where the kernel's principal role is the \textbf{inter-agent orchestration} of resources. Systems like AIOS~\citep{mei2025aios} and KAOS~\citep{zhuo2024kaos} exemplify this approach. A useful analogy is to view these frameworks as the agentic equivalent of \textbf{Kubernetes}: they orchestrate a fleet of agent-containers but do not govern the logic \emph{inside} them.

This `Kernel-as-Scheduler' model remains highly relevant and is a core component of Microsoft's Agent Framework, which directly inherits this capability from its AutoGen heritage~\citep{wu2023autogenenablingnextgenllm}. However, as with its predecessors, this model focuses on the orchestration \textit{between} agents, treating the internal workings of each agent as a black box to be scheduled. It does not architecturally address the fine-grained, internal execution safety of the agent process itself.

\subsubsection{Application-Level Governance Frameworks}

A second approach embeds constitutional checks directly within the agent's own logic. In this model, governance is a convention, not an architectural guarantee; it relies on the agent's probabilistic logic to follow rules, which can be subverted. Frameworks like TrustAgent~\citep{hua2024trustagent} and multi-agent systems like CrewAI~\citep{CrewAIInc2025} operate at this level.

This paradigm of application-level governance finds its most modern and sophisticated expression in OpenAI's AgentKit. Its ``Guardrail'' and ``Approval'' nodes are, in effect, constitutional checks that a developer must explicitly drag and drop into the agent's visual workflow~\citep{openai2025agentkit}. While offering significant power and flexibility, this confirms the core nature of the paradigm: governance is a feature of the application's logic, not an intrinsic, guaranteed property of the underlying execution runtime. Its correct implementation remains the responsibility of the developer.

To crystallize these architectural distinctions, Table~\ref{tab:governance_comparison} provides a direct comparison of the governance mechanisms in leading platforms against the paradigm proposed by ArbiterOS. As illustrated, current approaches, while powerful, embed governance as a configurable \textit{feature} of the application's logic, placing the burden of enforcement on the developer. This highlights the architectural gap that ArbiterOS is designed to fill: elevating governance from a feature to an intrinsic, guaranteed property of the underlying runtime.

\subsubsection{The `Kernel-as-Governor': Intra-Agent Governance}

ArbiterOS is designed to fill this specific architectural gap by introducing a complementary `Kernel-as-Governor' paradigm. Its focus is not on inter-agent orchestration, but on \textbf{intra-agent governance}. 

To extend the analogy, if a `Kernel-as-Scheduler' is Kubernetes, then ArbiterOS is analogous to a \emph{managed, high-assurance runtime like the JVM or .NET CLR}. It operates \textit{inside} the agent process, providing the structured execution environment and safety guarantees necessary for the agent's logic to run reliably. This governance-first approach is made possible by two key architectural innovations: the Agent Constitution Framework (ACF), a formal governance ISA, and the Hardware Abstraction Layer (HAL), which manages the `non-stationary hardware' of the underlying Probabilistic CPU.

\subsection{Connections to Foundational Disciplines}

The ArbiterOS paradigm synthesizes and extends principles from several foundational academic disciplines. Positioning our work in this broader context is essential for clarifying its contributions and intellectual lineage.


\subsubsection{Cognitive Architectures and Neuro-Symbolic Systems}

The design of the Agent Constitution Framework (ACF) is a form of cognitive architecture. It shares motivations with conceptual frameworks like CoALA (Cognitive Architectures for Language Agents), which also proposes organizing agents with modular memory components, a structured action space, and a generalized decision-making process~\citep{sumers2023cognitive}. However, where CoALA provides a valuable descriptive ``blueprint" for describing agents, ArbiterOS provides an enforceable architecture and engineering discipline for building them reliably.

Furthermore, the neuro-symbolic design of ArbiterOS aligns with a growing body of research advocating for the fusion of LLMs with formal methods to achieve trustworthiness~\citep{zhang2024fusionlargelanguagemodels,d2020neurosymbolic}. Both recognize the necessity of combining probabilistic and deterministic components to mitigate the unreliability inherent in purely learning-based systems. The difference lies in scope and application. Formal methods research often focuses on specific techniques for proving narrow properties, a task that remains challenging for large-scale neural networks. ArbiterOS, however, uses the neuro-symbolic split as the foundation for a comprehensive engineering paradigm, including the EDLC and abstractions for organizational scalability. It integrates formal methods as a practical tool within its Gradient of Verification, allowing developers to apply high-rigor checks where the Reliability Budget justifies the cost.

\subsubsection{Deterministic Workflow Orchestration}

While ArbiterOS employs a graph-based execution model, it is fundamentally distinct from traditional workflow orchestrators like Apache Airflow or Temporal. These systems were designed to ensure the reliable completion of deterministic tasks, where failure is an exception to be retried. ArbiterOS, in contrast, is designed to govern an inherently probabilistic substrate where non-determinism and error are expected operational characteristics. Its purpose is not merely to schedule tasks but to provide semantic, policy-driven governance over an agent's internal reasoning process, a capability for which traditional orchestrators have no analogue.

\subsubsection{Language Model Programming (LMP) Disciplines}

The ArbiterOS paradigm is complementary to, but distinct from, the emerging discipline of Language Model Programming (LMP), as exemplified by frameworks like DSPy~\citep{khattab2023dspycompilingdeclarativelanguage}. While both seek to replace brittle, hand-tuned prompting with a more systematic approach, their optimization targets are fundamentally different, creating a powerful ``micro vs. macro'' distinction.

\begin{itemize}
    \item \textbf{LMP optimizes the micro-architecture of a single instruction.} An LMP ``compiler'' like DSPy tunes prompts, few-shot examples, and model choices to maximize a specific task performance metric (e.g., answer accuracy on a given dataset). Its goal is to produce the \textit{best possible output}.

    \item \textbf{ArbiterOS governs the macro-architecture of the entire agent workflow.} An ArbiterOS ``Agent Compiler'' (as envisioned in Section~\ref{subsec:compilers}) would optimize the structural properties of the execution graph for system-level metrics like reliability, security, and latency. Its goal is to ensure the \textit{process is safe and efficient}.
\end{itemize}

The two paradigms are therefore highly synergistic. A developer could use DSPy to optimize the prompt template for a specific \textsc{Generate} Instruction Binding. ArbiterOS would then govern the execution of that performance-optimized instruction, ensuring it operates within the architecturally enforced safety and reliability guarantees of the broader agent system.

\subsection{Why Model Capabilities Cannot Replace Architectural Guarantees}

A forward-looking, ``model-centric'' position holds that any external OS-like structure will eventually be rendered obsolete by the sheer capability of future foundation models. This hypothesis suggests that a sufficiently advanced model could internalize its own safety, reliability, and verification logic, making an external governor unnecessary.

While models will undoubtedly become astoundingly more capable, this perspective conflates two fundamentally distinct categories of system properties:

\begin{itemize}
    \item \textbf{Model-Intrinsic Properties} concern the quality of reasoning and generation. A future model may achieve near-perfect accuracy in generating safe code or verifying facts. These are, however, inherently \textit{probabilistic} assurances about cognitive performance.
    \item \textbf{Architectural Guarantees} concern the deterministic invariants of the runtime system. They must hold true regardless of the behavior of any single component, and they are non-negotiable in high-stakes environments.
\end{itemize}

Consider the concrete engineering realities that demand architectural guarantees---realities that exist outside the model's cognitive process:

\begin{enumerate}
    \item \textbf{Managing Systemic Failures:} An agent must operate in an imperfect world. An external API will return a `503 Service Unavailable` error, a network connection will time out, and incoming data will be corrupted. These are not reasoning failures for the model to solve; they are systemic runtime exceptions that demand guaranteed, deterministic error handling, such as invoking a \textsc{Fallback} plan.

    \item \textbf{Enforcing Hard Resource Constraints:} A mission-critical system operates under strict, non-negotiable budgets. An agent cannot be politely asked to stay under a \$5 token limit or a 500ms latency budget. These are hard constraints that must be enforced by an external \textsc{Monitor\_Resources} check that deterministically terminates or reroutes execution.

    \item \textbf{Ensuring Auditable Compliance:} For any agent operating in a regulated domain (e.g., finance and healthcare), a probabilistic assurance 
    is insufficient. The business requires an immutable, cryptographic-grade audit trail---a ``Flight Data Recorder'' trace---that \textit{proves} a specific \textsc{Verify} step was executed. This is a systemic requirement for traceability, not a task the model can perform on its own.
\end{enumerate}

\begin{figure*}[!t]
\centering
    \includegraphics[width=0.95\linewidth]{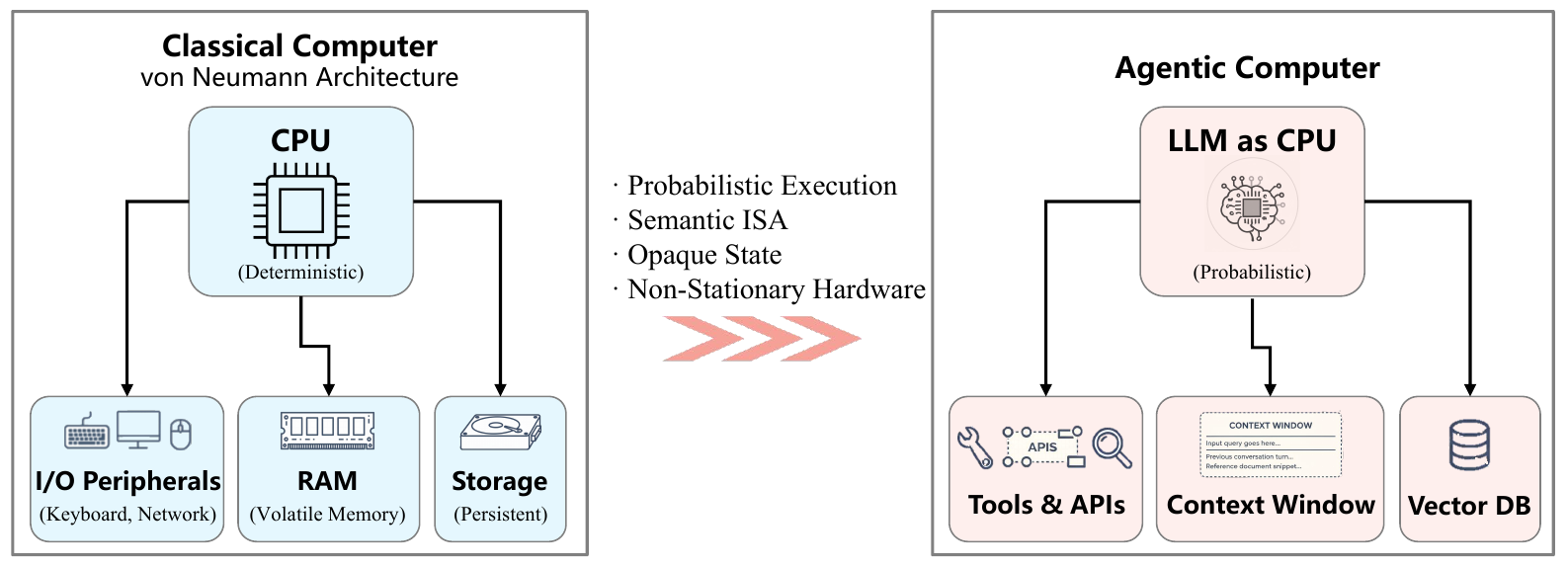}
    \caption{A side-by-side architectural comparison of a classical von Neumann computer and the proposed Agentic Computer, mapping core components to their agentic analogues and highlighting the fundamental shift from deterministic to probabilistic processing.}
    \label{fig:analogy}
\end{figure*}

This distinction is not philosophical; it is a fundamental principle of robust system design: \textbf{the governor must be separate from the governed}. Expecting a model to enforce these systemic rules on itself is a category error. It is akin to allowing a brilliant lawyer to write a contract and also serve as the sole judge and jury in any dispute over it. The model can be the brilliant lawyer that drafts the agent's logic, but it cannot be its own final, deterministic arbiter. This demonstrates the enduring and non-negotiable need for an external, architectural governance layer like ArbiterOS to safely manage an agent's interaction with the world.

\subsection{The Next Architectural Frontier: From Knowledge to Governed Experience}
\label{sec:knowledge_vs_experience}

While the ``crisis of craft" we identify is rooted in the challenge of managing today's \textit{knowledge-centric}, LLM-based systems, a broader debate within the AI community shapes the motivation for a new architectural paradigm. This debate concerns the fundamental source of intelligence and exposes a critical architectural gap that ArbiterOS is designed to fill.

On one hand, current agentic systems are quintessential artifacts of the knowledge-centric approach. Their underlying LLMs (the \textit{Probabilistic CPU}) are pre-trained on vast, static corpora of human-generated knowledge. While this provides them with impressive, broad capabilities, it also renders them brittle, non-adaptive, and ungrounded in real-world consequences, leading directly to the reliability challenges discussed.

On the other hand, a compelling forward-looking vision, powerfully articulated in~\cite{Sutton2025} argues for an ``Era of Experience." This \textit{experience-centric} approach posits that true, robust intelligence will not emerge from systems that merely mimic human knowledge, but from autonomous agents that learn continually by interacting with their environment. Such agents would build and refine their own models of the world based on direct feedback, rather than relying on a static, pre-trained understanding.

This creates a profound architectural dilemma. The knowledge-centric approach has hit a reliability wall, demanding a more robust engineering discipline. Yet, the experience-centric alternative, which involves deploying autonomous, self-modifying agents, presents immense and unresolved challenges in safety, alignment, and control. Unleashing a system that can learn and change its own behavior without a rigorous governance framework is an unacceptable risk.

This is precisely the architectural tension ArbiterOS is designed to resolve. Its governance-first paradigm is not an argument for one approach over the other, but rather a necessary substrate for both.
\begin{itemize}
    \item \textbf{For today's knowledge-centric agents}, the \textit{Symbolic Governor} provides the deterministic, architectural guarantees needed to manage the inherent unreliability of the \textit{Probabilistic CPU}, solving the immediate reliability gap.
    \item \textbf{For tomorrow's experience-centric agents}, this same architecture provides the essential safety framework. The \textit{Symbolic Governor} acts as a trusted, auditable supervisor, allowing a learning agent to safely explore, adapt, and self-modify within the bounds of human-defined constitutional rules.
\end{itemize}

Therefore, the motivation for ArbiterOS is twofold. It is an immediate and practical solution to the engineering crisis facing current agent development, and it is a foundational architectural paradigm providing the necessary safety and control for the future of governed, autonomous learning.

%% file: content/Sec2.tex
\section{The Agentic Computer}
\label{sec:computer&budget}

To move from craft to an engineering discipline, we must first adopt a mental model that accurately reflects the unique nature of the hardware we are commanding. This paper introduces the \textbf{Agentic Computer}, a formal analogy that maps the components of an LLM-based agent to the classical von Neumann architecture~\citep{eigenmann1998neumann}. The parallels are intuitive—the LLM serves as the \textbf{Probabilistic Central Processing Unit (CPU)}, the context window as fast, volatile \textbf{RAM}, and external tools as \textbf{I/O Peripherals}, as shown in Fig.~\ref{fig:analogy}.

However, the true utility of this model lies in highlighting their stark divergences. These differences are not flaws to be eliminated; they are the fundamental, unchangeable properties of this new `hardware' that any robust system must be designed to handle. They define the core engineering challenge of the agentic era. We identify five such properties:

\begin{itemize}
    \item \textbf{Probabilistic Execution:} Unlike a silicon CPU, a Probabilistic CPU is non-deterministic. The same input can yield different, often equally valid, outputs. Logical errors and hallucinations are not rare exceptions but expected operational characteristics of the hardware~\citep{huang2025survey}. This inherent unreliability demands that error handling and recovery paths be treated as core, first-class architectural primitives.
    \item \textbf{Semantic Instruction Set:} A classical CPU executes unambiguous binary opcodes. In contrast, the Probabilistic CPU interprets the semantic intent of natural language. Its Instruction Set Architecture (ISA) is therefore unstable and exquisitely sensitive to phrasing and tone, a problem that cannot be solved by simply finding the ``perfect prompt"~\citep{Lu2021FantasticallyOP}. This instability necessitates a formal, sanitizing boundary between the model's ambiguous intent and the deterministic execution of high-stakes tools.
    \item \textbf{Opaque Internal State:} The internal ``thought process" of an LLM is an uninterpretable, high-dimensional vector state—an opaque ``black box"~\citep{belinkov2019analysis}. This lack of inspectability presents a profound challenge for debugging and alignment, making process-level transparency—an explicit, auditable trace of every observable state change and action—a non-negotiable architectural requirement for building trust.
    \item \textbf{Non-Stationary Hardware:} A silicon CPU's ISA is stable for decades. The underlying Probabilistic CPU—the LLM itself—is constantly evolving, with each new model release being equivalent to a hardware replacement. This constant evolution renders any static solution brittle by design. It demands both a development discipline centered on continuous re-verification and, critically, a formal \textbf{Hardware Abstraction Layer (HAL)} that decouples the agent's durable logic from the volatile, model-specific implementation.
    \newpage
    \item \textbf{Volatile and Unreliable Memory:} The context window, which serves as the Agentic Computer's RAM, is not a stable storage device. Its unique properties transform memory management from a simple optimization into a core reliability challenge. Unlike the deterministic memory of a classical computer, it suffers from three key forms of unreliability:
    \begin{itemize}
        \item \textbf{Semantic Eviction:} Instead of a predictable rule like `Least Recently Used,' the context window forgets information based on a probabilistic assessment of `relevance.' This eviction is a cognitive task, not a deterministic one, and is prone to accidentally discarding critical data.
    
        \item \textbf{Attention Variance:} The location of information within the context window matters. Due to effects like the ``lost-in-the-middle'' problem, data placed in the center of the context can be ignored or given less weight, making memory placement a critical, non-neutral decision~\citep{Liu2023LostIT}.
    
        \item \textbf{Lack of Virtual Memory:} Once information is evicted from context, it is gone completely. There is no automatic paging system to retrieve it from a backing store; it must be explicitly and consciously reloaded into memory.
    \end{itemize}
    
    The unreliability of this memory system creates a dangerous trap. Common context management techniques, such as using the LLM itself to summarize past conversation, are themselves high-risk cognitive operations. When developers treat these operations as if they were safe, deterministic functions, they risk the silent corruption of the agent's internal state. We term this insidious failure mode \textbf{Cognitive Corruption}: the unobserved degradation of an agent's working memory. This corruption manifests in two ways: \textbf{Errors of Omission}, where the agent forgets or ignores critical facts due to semantic eviction or attention variance; and \textbf{Errors of Commission}, where the agent's memory becomes polluted with hallucinations introduced by a flawed summarization. This degradation leads to unpredictable and catastrophic downstream failures and serves as a primary motivation for why context management cannot be left to ad-hoc application logic and fragile prompt chains~\citep{mei2025surveycontextengineeringlarge}. It must be a fundamental, governed responsibility of the OS.
\end{itemize}

Taken together, these five properties create a fundamental economic challenge. Every layer of governance needed to manage this volatile hardware—every verification step, fallback plan, and abstraction layer—imposes an ``Abstraction Tax" in latency, computation, and complexity. To make this trade-off explicit, we introduce the \textbf{Reliability Budget}: the total investment a project is willing to make to ensure a safe and successful outcome, a value dictated by the cost of failure. The central failure of the artisanal approach is its inability to manage this budget predictably, leading to systems that are either too expensive to run or too brittle to trust.


\begin{figure*}[t!]
    \centering
    \includegraphics[width=0.85\linewidth]{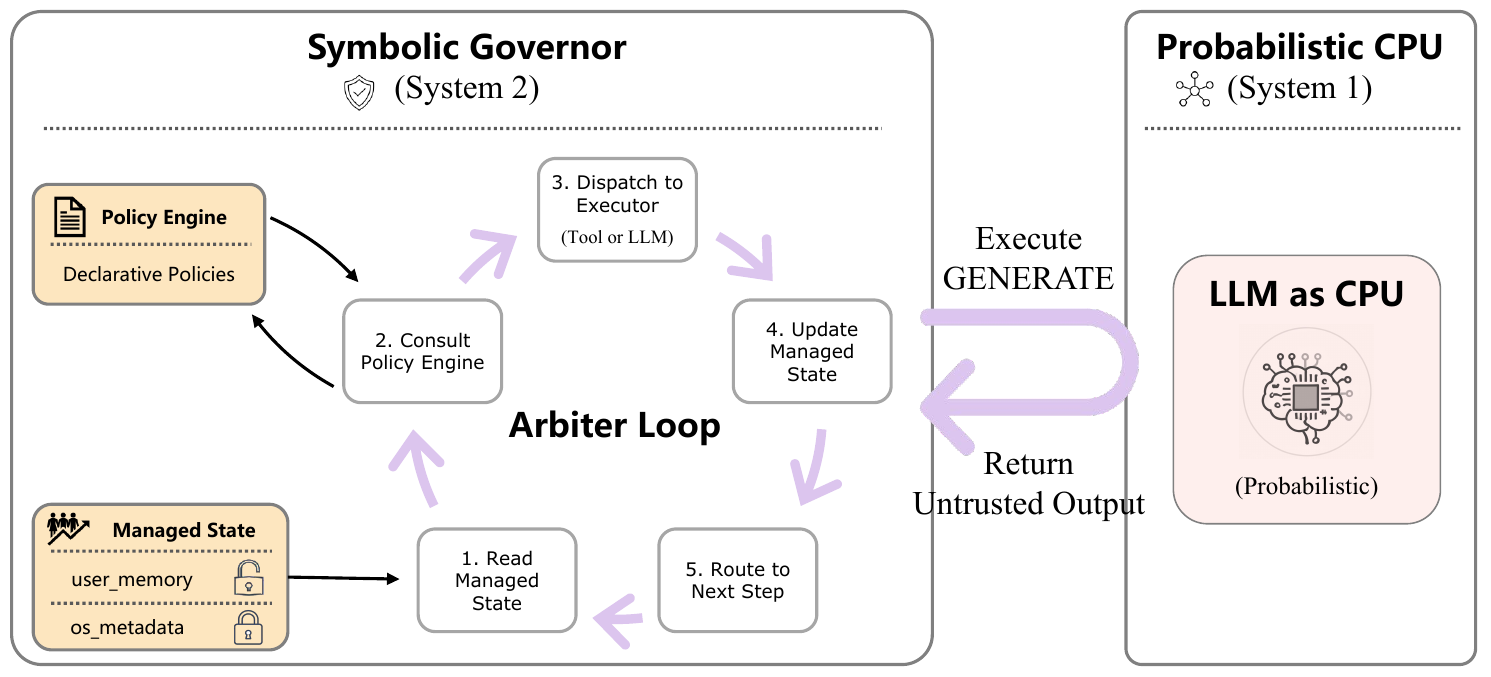}
    \caption{The ArbiterOS neuro-symbolic architecture. The deterministic Symbolic Governor (Kernel) orchestrates the agent's workflow, managing the trusted state and enforcing policies. It governs the probabilistic CPU (LLM) through a formal instruction set, making decisions via the Arbiter Loop, a non-bypassable kernel function that intercepts and validates every state transition before executing the next instruction.}
    \label{fig:neuralSymbolicArc}
\end{figure*}

A budget is useless without a mechanism to spend it. We therefore introduce the \textbf{Gradient of Verification} as the primary framework for investing the Reliability Budget. This gradient transforms reliability from an abstract goal into a concrete architectural choice, ranging from flexible but probabilistic checks (e.g., using an LLM-as-Judge~\citep{zheng2023judging}) to rigorous checks with formal logic (see Appendix~\ref{sec:gradverif}). This makes the cost of reliability a measurable and justifiable engineering decision.

Furthermore, ArbiterOS provides systemic patterns for managing the residual risk inherent in probabilistic checks. For instance, the Arbiter Loop can be configured to use \textbf{Ensemble Verification}, deriving a more robust signal from a consensus of multiple diverse LLM-judges. It can also perform \textbf{Confidence-Based Escalation}, a policy where a low-confidence score from a probabilistic check (e.g., `$p < 0.8$') deterministically triggers a higher-rigor check or a mandatory human review. These patterns demonstrate how the OS can transform uncertain signals into architecturally sound risk management strategies.

This reframing leads to a stark but essential conclusion. The fundamental properties of the Agentic Computer, coupled with the need to consciously manage a Reliability Budget, reveal that a simple, monolithic agent loop is architecturally insufficient. The task of managing a processor with an unstable ISA, an opaque state, high-risk memory, and an expected high fault rate demands a higher-level abstraction that provides systemic services for process management, error handling, and resource governance. This is the formal, time-tested role of an operating system.


%% file: content/Sec3.tex
\section{ArbiterOS: A Neuro-Symbolic Operating System Paradigm}
\label{sec:arbiteros}


The preceding analysis of the agentic ecosystem reveals a critical architectural gap. While the \textbf{`Kernel-as-Scheduler'} paradigm addresses orchestration \textit{between} agents and \textbf{`Application-Level Governance'} embeds safety as a configurable \textit{feature}, neither provides a trusted, architecturally-enforced runtime to govern the inherently unreliable workflow \textit{within} a single agent.

To fill this specific gap in \textbf{intra-agent governance}, this paper introduces \textbf{ArbiterOS}. It is a formal operating system paradigm built on the complementary principle of the \textbf{`Kernel-as-Governor'}. This architecture is designed not for traditional software processes, but specifically to command the unique hardware of the \textbf{Agentic Computer}. Its design is a practical neuro-symbolic architecture~\citep{wan2024cognitiveaisystemssurvey}, crafted to harness the power of the Probabilistic CPU while rigorously mitigating its inherent unreliability through a deterministic, trusted Governor.

This architecture achieves its goal by separating the agentic system into two distinct components (see Fig.~\ref{fig:neuralSymbolicArc}). This division of labor is analogous to the ``System 1" (fast, intuitive) and ``System 2" (slow, deliberate) models of cognition~\citep{Kahneman2011}. While LLMs can be prompted to exhibit both types of reasoning~\citep{li202512surveyreasoning}, the foundational split in ArbiterOS is more architecturally precise: it is the immutable separation between the untrusted, probabilistic reasoning engine and the deterministic, trusted governor that controls the reasoning process.

\begin{itemize}
    \item \textbf{The Probabilistic CPU (``System 1"):} This is the agent's \textbf{`neural' component}---the Large Language Model itself. It excels at processing unstructured data through associative, heuristic-driven reasoning, making it powerful for tasks like understanding natural language, generating creative plans, and synthesizing complex information. By architectural design, its outputs cannot be fully trusted and must be subjected to external verification by the Symbolic Governor.

    \item \textbf{The Symbolic Governor (``System 2"):} In direct contrast, this is the agent's deterministic, auditable \textbf{`symbolic' component}---the ArbiterOS Kernel. It is a rule-based engine responsible for the high-level orchestration of the agent, the strict enforcement of declarative policies, and making discrete, verifiable decisions. It acts as the system's arbiter of trust, governing the Probabilistic CPU through a formal instruction set.
\end{itemize}


This `Kernel-as-Governor' paradigm is architecturally distinct from recent industry approaches. The ArbiterOS kernel is not a component \textit{in} the agent's workflow; it is the trusted, deterministic runtime \textit{that executes} the workflow, enforcing policies defined in the formal Agent Constitution Framework (ACF). This separation is operationalized through a set of core OS primitives: 


\begin{itemize}
    \item \textbf{The Managed State:} The foundation of all governance is the \textbf{Managed State}, the OS's central and serializable source of truth. Its formal separation of \texttt{user\_memory} from protected \texttt{os\_metadata} is the prerequisite for auditable logging and fault tolerance. While new frameworks provide excellent observability tools like AgentKit's ``trace grading'' or Microsoft's deep OpenTelemetry integration, the `Managed State' is the OS primitive that elevates observability to true, kernel-enforced \textbf{reproducibility} and high-fidelity ``time-travel'' debugging.

    \item \textbf{The Arbiter Loop:} 
    This is the heart of the operating system and the core of its scheduler, providing the fundamental guarantee of \textit{Process-Level Determinism}. The ArbiterOS runtime does not permit direct transitions between instructions in the agent's Execution Graph. Instead, after any instruction is executed, the scheduler deterministically yields control and the current \textsc{Managed State} to the Arbiter Loop. This non-bypassable interception is the foundational mechanism that allows the Governor to operate. Only after the Arbiter Loop inspects the protected \texttt{os\_metadata}, validates the state transition against the active policies, and makes a trusted routing decision is the next instruction allowed to execute. 
    Its ability to transform uncertain probabilistic signals (e.g., a low confidence score from an LLM-as-judge) into certain, auditable \emph{actions} (e.g., triggering a human review via \texttt{INTERRUPT}) is the core of the Governor's power, enabling the system to safely leverage heuristic checks without compromising its architectural trustworthiness.
\end{itemize}

\subsection{Managing Governance: From Risk Profiles to Performance Tiers}
\label{subsec:managing_governance}

The true power of this paradigm lies in shifting the developer's task from imperatively coding every check to declaratively defining the required level of governance via a \textbf{Policy Engine}. This configurability is the key to managing both risk and performance. Policies are grouped into \textbf{Configurable Execution Environments}, allowing governance to be tuned for both an agent's specific risk profile and its current lifecycle stage.

\paragraph{Expressiveness of the Policy Language.}
While the stateless transition policies discussed in this paper (e.g., enforcing a ``think then verify'' workflow) are foundational, we acknowledge that a production-grade implementation requires a more expressive policy language. The ArbiterOS \texttt{Policy Engine} is therefore envisioned to evolve beyond simple transition validation to support more dynamic, context-aware rules. Future work would focus on enabling:
\begin{itemize}
    \item \textbf{Stateful Policies,} which make decisions based on the agent's ``Managed State'' (e.g., ``forbid a \texttt{TOOL\_CALL} to a payment API if \texttt{user\_memory.has\_flag('high\_risk\_user')}'').
    \item \textbf{Temporal Policies,} for enforcing constraints like rate-limiting over time (e.g., ``permit a \texttt{TOOL\_CALL} to this specific API no more than once every 5 seconds'').
    \item \textbf{Resource-Based Policies,} which perform more complex checks against budgets (e.g., ``only allow a high-cost \texttt{DECOMPOSE} step if the remaining \textit{Reliability Budget} is above 50\%'').
\end{itemize}
This evolution transforms the Policy Engine from a static architectural linter into a true, dynamic governance runtime capable of enforcing sophisticated operational constraints.

\paragraph{Configurability in Practice.}
For managing risk, an \texttt{Executor} environment can load policies that enforce a strict ``think then verify'' workflow, whereas a \texttt{Strategist} environment may use policies that prioritize metacognitive checks to avoid wasted resources (see Appendix~\ref{app:sec:policies}). For managing performance, developers can define tiers of governance:
\begin{itemize}
    \item \textbf{Development Environment:} Optimized for rapid iteration, disabling costly checks to focus on logic development while still capturing a ``Flight Data Recorder'' trace for debugging.
    \item \textbf{Staging Environment:} Optimized for fidelity, enabling the full suite of verification checks to catch regressions against the Golden Dataset.
    \item \textbf{Production Environment:} Optimized for low-latency performance, running a pre-validated or ``compiled'' version of the policies to minimize overhead.
\end{itemize}

Furthermore, not all checks must be synchronous. For non-blocking validations, ArbiterOS supports \textbf{asynchronous verification}. The Arbiter Loop can dispatch a low-priority check (e.g., on stylistic tone) to a background process, allowing the main workflow to proceed without added latency while still flagging violations for logging. This tiered and asynchronous approach transforms governance from a rigid overhead into a flexible, performance-aware engineering discipline. It provides the framework for managing the inherent costs of a structured architecture, which we formally define as the \textbf{``Abstraction Tax.''}

\subsection{The Abstraction Tax as a Strategic Investment}

The ``Abstraction Tax'' introduced by this formal paradigm has two components: \textbf{Performance Overhead} (latency, compute) and \textbf{Engineering Overhead} (developer complexity). As we have shown, this tax is not a fixed penalty but a manageable and optimizable variable. To make it tangible, consider a synchronous API call in a high-reliability production environment. A naive agent might take 400ms. A governed agent might first execute a deterministic \texttt{VERIFY} instruction (50ms) and log the result (10ms), introducing a 15\% latency tax. This is the explicit cost paid to prevent an entire class of silent failures.

A core thesis of this paper is that this tax is a strategic investment that yields returns in reliability and, counter-intuitively, in efficiency. The same governance primitives that ensure reliability can also provide performance gains: governed memory management reduces expensive token consumption, while metacognitive oversight prevents wasted computation.

This formal structure is also the prerequisite for systematically \textit{reducing} the tax over time. The `Agent Compiler' vision (Section~\ref{subsec:compilers}) targets the automated reduction of Performance Overhead, while a dedicated `Tooling Ecosystem' (Section~\ref{subsec:cognitive_ide}) minimizes Engineering Overhead. Thus, the Abstraction Tax is the necessary foundation for building agents that are truly scalable, efficient, and maintainable.


Crucially, this investment need not be paid all at once. The ArbiterOS paradigm is designed around the core tenet of \textbf{Progressive Governance}, a philosophy that transforms the Abstraction Tax from a prohibitive upfront cost into a manageable, pay-as-you-go investment. The following section details this incremental adoption path, the primary mechanism for making principled agent engineering practical and accessible.

\subsection{The Principle of Progressive Governance in Practice}
\label{subsec:progressive_governance}

As introduced above, the principle of Progressive Governance is the core tenet that makes the Abstraction Tax a manageable, strategic investment. It provides a practical, pay-as-you-go pathway from a brittle prototype to a robust, production-ready system, allowing teams to start with a ``Minimum Viable ArbiterOS'' and incrementally layer on more sophisticated guarantees as the system's reliability requirements increase.

This provides immediate value at every stage. A typical adoption journey follows three stages:

\begin{enumerate}
    \item \textbf{Stage 1: Gaining Auditability (The ``Flight Recorder'').} A team can begin by simply wrapping an existing, naive agent loop within the ArbiterOS runtime. With minimal code changes, this first step provides immediate and immense value: every step of the agent's execution is now logged as a deterministic, auditable trace within the \textsc{Managed State}. This ``time-travel debugging'' capability alone can drastically reduce the mean time to resolution (MTTR) for production failures.

    \item \textbf{Stage 2: Hardening for Resilience.} Once the agent is observable, the team can address the most common failure modes by introducing the first governance patterns. Typically, this involves adding a deterministic \texttt{Verify} instruction after a high-risk step (like an external API call) and registering a \texttt{Fallback} plan. This single architectural addition transforms fragile error handling into a guaranteed, resilient process.

    \item \textbf{Stage 3: Achieving Full Robustness.} Finally, as the agent matures, the team can layer on more advanced governance for internal cognitive failures and strategic oversight. This includes governed memory management (using \texttt{Compress}) to prevent Cognitive Corruption, or strategic self-correction loops (using \texttt{Evaluate\_Progress}) to prevent wasted computation.
\end{enumerate}

This tiered approach demonstrates how the Abstraction Tax is paid not as a lump sum, but as a series of targeted investments from the agent's Reliability Budget, each providing a concrete return in reliability and trustworthiness.

%% file: content/Sec4.tex
\section{The Agent Constitution Framework (ACF)}
\label{sec:acf}

\begin{figure*}[t!]
\centering
\includegraphics[width=0.85\linewidth]{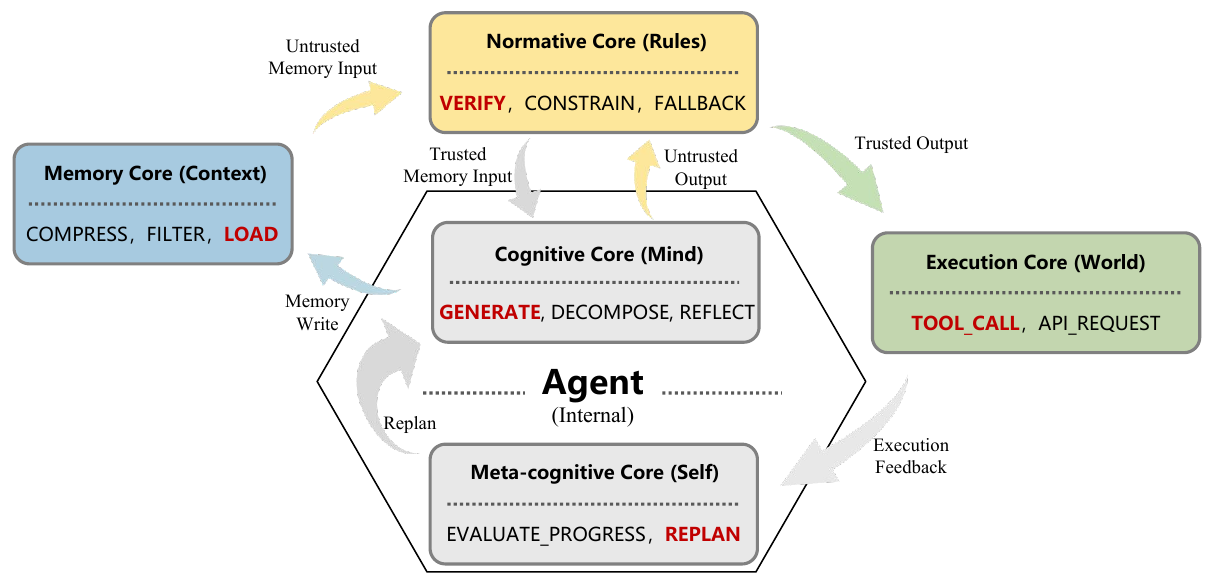}
\caption{The five operational cores of the Agent Constitution Framework (ACF). Instructions are categorized to create governable boundaries between the agent's internal cognitive world, its memory, its interactions with the external world, and the normative and metacognitive rules that constrain its behavior.}
\label{fig:acfDomains}
\end{figure*}

An operating system is useless without a formal language to command it. Just as a silicon CPU is governed by an Instruction Set Architecture (ISA), the Probabilistic CPU requires its own ISA to be programmed and governed effectively. We propose the \textbf{Agent Constitution Framework (ACF)} as this instruction set.

The crucial distinction is that the ACF is a \textbf{macro-architecture ISA designed purely for governance}, not a micro-architecture ISA for computation. The ArbiterOS kernel does not care \emph{how} an instruction performs its reasoning—that is the opaque, ``micro-architectural" domain of the Probabilistic CPU. Instead, the kernel only needs to know an instruction's formal type. This classification—for example, that a step is a probabilistic ``Cognitive" instruction whose outputs are untrusted—provides the unambiguous vocabulary for the Policy Engine to enforce architectural rules. It is this formal typing that transforms an agent's workflow from an opaque chain of thought into a governable, auditable process.


To enable this granular governance, the ACF is not a flat list of commands. Its structure is derived from the fundamental challenge of governing a thinking and acting entity. It organizes instructions into five distinct ``cores'' that establish the discrete, governable boundaries necessary for control. This decomposition is not arbitrary; it is derived from first principles by drawing the necessary architectural boundaries between an agent's internal reasoning (\textbf{Cognitive Core}), its working memory (\textbf{Memory Core}), its interactions with the external world (\textbf{Execution Core}), the human-defined rules it must obey (\textbf{Normative Core}), and its capacity for strategic self-reflection (\textbf{Metacognitive Core}). This structure separates the untrusted probabilistic reasoning from the trusted verification steps and high-stakes external interactions, providing the formal seams required for robust engineering.

The five cores are:

\begin{itemize}
    \item \textbf{The Cognitive Core (The Mind):} Manages the internal, probabilistic reasoning of the Probabilistic CPU. This includes foundational instructions for generating novel content (\texttt{GENERATE}), creating structured plans (\texttt{DECOMPOSE}), and performing self-critique (\texttt{REFLECT}). Its outputs are fundamentally untrusted and form the primary subject of governance by the other cores.
    
    \item \textbf{The Memory Core (The Context):} Governs the agent's working memory. This core formalizes high-risk cognitive operations on the agent's state, such as summarization (\texttt{COMPRESS}), selective recall (\texttt{FILTER}), and external data integration (\texttt{LOAD}). It provides the primitives to manage the Probabilistic CPU's limited context window not merely as a storage space, but as a governable resource, preventing the \textbf{Cognitive Corruption} introduced in Section~\ref{sec:computer&budget}.
    
    \item \textbf{The Execution Core (The World):} Provides the interface to the external, deterministic world. It contains instructions for interacting with tools and APIs (\texttt{TOOL\_CALL}) to enact decisions or retrieve data. Actions in this core are often high-stakes and must be preceded by rigorous checks from the Normative Core to ensure safety.
    
    \item \textbf{The Normative Core (The Rules):} Enforces the human-defined rules, policies, and safety constraints. This core provides the critical primitives for reliability, such as checking outputs for correctness (\texttt{VERIFY}), ensuring compliance with ethical guidelines (\texttt{CONSTRAIN}), and executing resilient recovery plans (\texttt{FALLBACK}). It acts as the system's deterministic arbiter, governing the transition from untrusted thought to trusted action.
    
    \item \textbf{The Metacognitive Core (The Self):} Enables strategic oversight of the agent's own performance. Instructions in this core, such as \texttt{EVALUATE\_PROGRESS}, allow the agent to detect and escape unproductive reasoning paths or ``rabbit holes," ensuring the efficient use of computational resources in complex, long-horizon tasks.
    \end{itemize}

This architecture, organized by cores, provides the discrete, governable boundaries necessary for control. For example, the ArbiterOS kernel can enforce a policy mandating that a probabilistic \texttt{GENERATE} instruction (Cognitive Core) must be followed by a deterministic \texttt{VERIFY} instruction (Normative Core) before a high-stakes \texttt{TOOL\_CALL} (Execution Core) is permitted. This ability to architecturally enforce a ``think then verify" workflow—a systemic, auditable safety guarantee—is impossible without such a formal, discrete instruction set. Instructions like \texttt{CONSTRAIN} also provide a practical runtime implementation for concepts like Constitutional AI~\citep{bai2022constitutionalaiharmlessnessai}, transforming abstract principles into enforceable rules. The complete ACF instruction set is detailed in Appendix~\ref{app:instset}.

To make these abstract instructions enforceable, each is connected to a concrete implementation through a formal \textbf{Instruction Binding}. This binding is a design-time, serializable contract that specifies an instruction's type, its implementation (e.g., a prompt template or a Python function), and strict, typed schemas for its inputs and outputs. This mechanism is the linchpin of the framework's security and portability model.

\begin{figure*}[t!]
    \centering
    \includegraphics[width=0.75\linewidth]{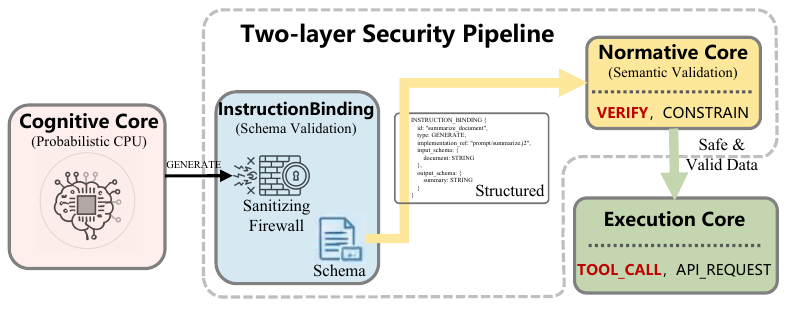}
    \caption{The ``sanitizing firewall" mechanism. An \textsc{InstructionBinding} enforces a strict output schema on the probabilistic \texttt{GENERATE} step. Valid, structured data is passed to the tool for execution, while malformed or malicious outputs (e.g., from a prompt injection attack) are blocked, preventing them from reaching the Execution Core.}
    \label{fig:acfFirewall}
\end{figure*}

First, the binding acts as a \textbf{sanitizing firewall}, providing a foundational layer of security. By enforcing schema validation on all outputs, it guarantees a critical principle: \textbf{LLMs produce structured data, not executable commands}. This architecture, enabled by robust structured output parsing libraries~\citep{567labs2023instructor}, fundamentally reduces the attack surface by eliminating the entire class of direct command injection attacks. However, this structural guarantee does not, by itself, prevent semantic manipulation (e.g., an attacker tricking the LLM into populating a valid JSON field with a malicious value). This is where the second layer of defense comes in. The \textbf{Normative Core} is responsible for semantic validation, using \texttt{VERIFY} and \texttt{CONSTRAIN} instructions to inspect the \emph{content} of the structured data for safety and correctness before it is used in high-stakes operations. Together, these two layers provide a robust, \textbf{defense-in-depth} security posture that is far superior to fragile, prompt-level defenses~\citep{chen-etal-2025-defense}.

Second, and just as critically, the Instruction Binding serves as the \textbf{`device driver'} for a given instruction on a specific Probabilistic CPU. The complete set of an agent's bindings thus functions as its \textbf{Hardware Abstraction Layer (HAL)}. This layer encapsulates all model-specific details—such as proprietary prompt templates or function-calling syntax—within the bindings themselves. This cleanly separates the \textbf{stable structure} of the Execution Graph from the \textbf{volatile, model-specific implementations}. This architectural decoupling is what \textbf{facilitates portability}. It transforms the high-risk process of migrating between foundation models from a system-wide rewrite into a manageable, localized engineering task focused on updating `drivers,' significantly reducing the cost and risk of vendor lock-in.

\paragraph{Extensibility of the ACF.} While the five foundational cores provide a universally applicable framework for governing a wide range of agentic systems, ArbiterOS is designed not as a dogmatic structure, but as an extensible one. The framework explicitly supports the definition of \textbf{Custom Cores and Instruction Bindings} to accommodate highly specialized, domain-specific tasks.

This allows developers to define new semantic categories of operations that are first-class citizens of the governance model. For example, an agent designed for complex algorithmic trading might implement a custom \texttt{QuantitativeCore} with specialized instructions like \texttt{EXECUTE\_BACKTEST} or \texttt{CALCULATE\_ALPHA}. These custom instructions, just like the foundational ones, are defined with formal types and schemas, making them fully governable by the Policy Engine. A policy could enforce that a high-compute \texttt{EXECUTE\_BACKTEST} instruction is always preceded by a \texttt{MONITOR\_RESOURCES} check to stay within budget, or that the output of \texttt{CALCULATE\_ALPHA} must pass a custom \texttt{VERIFY\_BOUNDS} check before influencing a trade. This capability for domain-specific extension ensures that ArbiterOS can be adapted from general-purpose assistants to highly specialized, mission-critical systems, all while maintaining a consistent and auditable governance layer.

%% file: content/Sec5.tex
\begin{figure*}[ht!]
    \centering
    \includegraphics[width=1.0\linewidth]{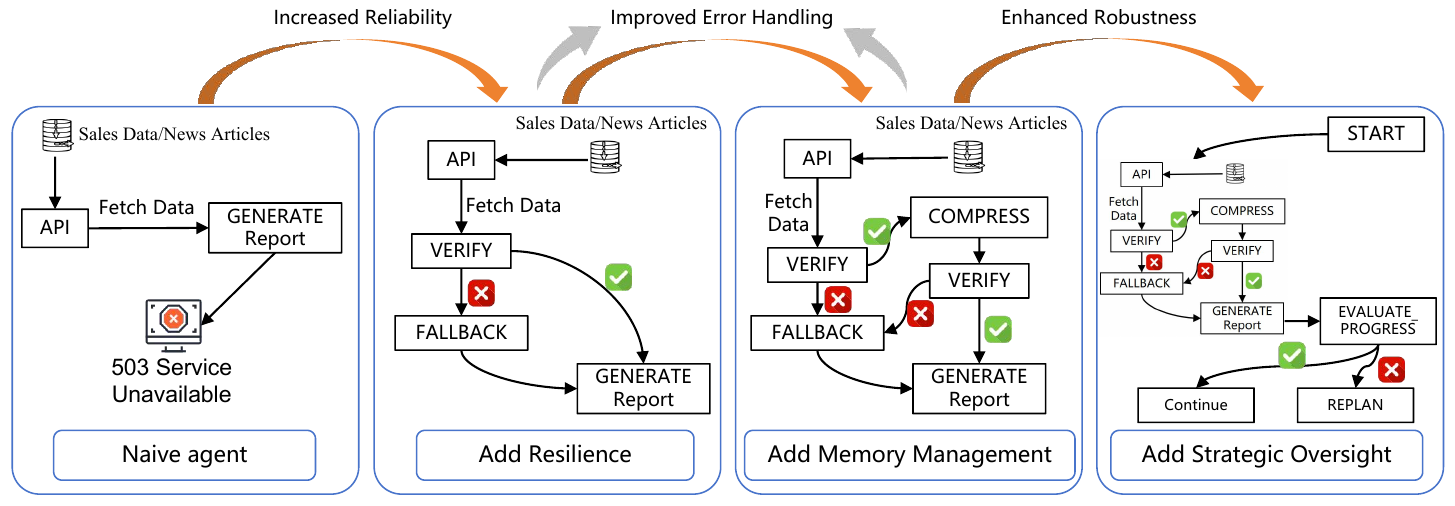}
    \caption{The architectural evolution of an agent under the Progressive Governance principle. The agent begins as a brittle prototype and is incrementally hardened by layering on formal governance primitives (\texttt{VERIFY}, \texttt{FALLBACK}), memory management (\texttt{COMPRESS}), and strategic oversight (\texttt{EVALUATE\_PROGRESS}) to create a robust, production-ready system.}
    \label{fig:archEvolution}
\end{figure*}

\section{Walkthrough: Implementing Progressive Governance in Practice}
\label{sec:governancePrinciple}


To demonstrate how the core ArbiterOS principle of \textbf{Progressive Governance} (introduced in Section~\ref{subsec:progressive_governance}) is applied in a real-world scenario, this section presents an illustrative walkthrough following its three-stage adoption journey. 

For this illustration, we use a common business task: generating a comprehensive market analysis report. This task requires a blend of agent capabilities, mapping directly to the Application Archetypes defined in Appendix~\ref{sec:taxonomyagent}: 

\begin{itemize}
    \item Synthesizing quantitative sales data (an \textsc{Executor} task);
    \item Summarizing qualitative news (a \textsc{Synthesizer} task);
    \item Analyzing the competitive landscape (a \textsc{Strategist} task).
\end{itemize}


Figure~\ref{fig:archEvolution} visually summarizes this architectural evolution. The panels illustrate the specific governance primitives that are layered onto the agent's execution graph. The ``Naive agent'' panel represents the brittle baseline. The ``Add Resilience'' panel corresponds directly to the architectural changes made in Stage 2 of our development journey. The final two panels, ``Add Memory Management'' and ``Add Strategic Oversight,'' provide two concrete examples of the advanced governance patterns that are implemented during Stage 3 to achieve full robustness. Note that Stage 1 of the journey---Gaining Auditability---focuses on observability and does not alter the agent's architecture itself, which is why it is not depicted as a separate flowchart.



\subsection{The Naive Prototype and Brittle Execution}


The initial implementation is a linear execution chain, for example:
\[
\texttt{GENERATE} \;\rightarrow\; \texttt{TOOL\_CALL} \;\rightarrow\; \texttt{GENERATE} ...,
\]
wherein the agent fetches financial data and assembles a report. When the financial API responds with an error (e.g., a \texttt{503 Service Unavailable}), the agent passes the raw HTML error string downstream, producing nonsensical claims such as \textit{“According to our sales data, there has been a ‘503 Service Unavailable’.”} This highlights the brittleness of ungoverned pipelines.

\subsection{Stage 1: Gaining Auditability (The ``Flight Recorder'')}



The first step in the Progressive Governance journey is to wrap the existing prototype within the ArbiterOS runtime. This minimal-effort action provides the immediate value of a deterministic \textbf{``Flight Data Recorder''} trace for every execution. Now, when the ``503 Service Unavailable'' error occurs, the team has a perfect, step-by-step record of the failure. This trace transforms debugging from a speculative art into a deterministic science, providing the essential diagnostic data needed for the next stage.

\subsection{Stage 2: Hardening for Resilience}

With clear diagnostics from Stage 1, we now address this brittleness by introducing primitives from the \textbf{Normative Core}. The \texttt{VERIFY} instruction establishes a deterministic checkpoint to validate that API responses adhere to a formal schema, and the \texttt{FALLBACK} instruction encodes a pre-defined recovery pathway to a cached data source. The formal definitions for these primitives are specified as \textbf{Instruction Bindings}, the serializable contracts that make governance enforceable (see Appendix~\ref{app:sec:walkthrough_details} for pseudocode examples).

The agent's ability to gracefully degrade is captured in its execution trace. When the primary API fails, the Arbiter Loop deterministically reroutes execution to the \texttt{FALLBACK} plan, as detailed in the execution trace in Appendix~\ref{app:sec:walkthrough_details} (Table~\ref{tab:api-outage}). This architectural enforcement transforms brittle error handling into a governed, auditable process capable of graceful degradation.

\subsection{Stage 3: Achieving Full Robustness}


Having secured the agent against external failures, the team proceeds to Stage 3: layering on advanced governance for the agent's internal cognitive and strategic processes.

First, to combat \textbf{Cognitive Corruption}, the developer introduces a \emph{governed memory pattern}. This involves layering a safety check onto the high-risk \texttt{COMPRESS} instruction. As deterministically verifying the fidelity of a summary is often impossible, the developer instead uses a probabilistic, \textbf{Level 1} check: another LLM is tasked to act as a judge, comparing the summary against the original text to check for factual consistency and the preservation of critical details. 

The ArbiterOS Policy Engine is then configured to manage the residual risk of this probabilistic check. The policy might stipulate that if the LLM-as-judge returns a high confidence score (e.g., `$p > 0.9$'), the process continues. However, if the confidence is low, the policy uses \textbf{Confidence-Based Escalation} to deterministically trigger an \texttt{INTERRUPT} instruction, pausing the agent and flagging the summary for mandatory human review.

This governed pattern produces a resilient, efficient, and risk-aware memory process. The concise summary is retained in the context window, ensuring the final report is both coherent and factually complete, all while providing an architectural safety net for a fundamentally non-deterministic operation. This not only solves the reliability issue but also \textbf{boosts performance and reduces cost} by significantly lowering the token count on subsequent steps.


Second, to extend robustness from operational tasks to strategic reasoning, the developer hardens the agent against unproductive ``rabbit holes'', returning irrelevant news after a broad web search. 
To mitigate this, an \textsc{Evaluate\_Progress} instruction from the \textbf{Metacognitive Core} is introduced. If this check returns a `FAIL' signal, the Arbiter Loop deterministically routes execution to a \textsc{Replan} step. The agent's flawed reasoning is thus caught by the OS itself, allowing it to self-correct its strategy. This strategic self-correction is captured in the execution trace in Appendix~D.3 (Table~10), demonstrating how governance is now applied not just to an instruction's output, but to the strategic validity of the agent's own reasoning process.


With these additions, the agent is not only resilient to external errors but is also robust against its own internal cognitive and strategic failures.

\subsection{Conclusion of the Walkthrough}


This walkthrough has demonstrated the principle of Progressive Governance in action. The agent evolved systematically through a practical, repeatable workflow: first gaining \textbf{auditability} to diagnose failures deterministically, then achieving \textbf{resilience} against external errors, and finally attaining full \textbf{robustness} against its own internal failure modes. This journey shows how the ``Abstraction Tax'' is not a monolithic barrier, but a series of manageable, value-driven investments that transform a brittle prototype into a trustworthy, production-ready system.

%% file: content/Sec6.tex
\section{The Discipline: The Evaluation-Driven Development Lifecycle (EDLC)}
\label{sec:discipline}


A new architectural paradigm demands a new development discipline. Traditional software lifecycles are fundamentally ill-suited for the empirical and probabilistic nature of agentic engineering. Without a structured approach to evaluation, reliability becomes a matter of guesswork and progress is difficult to measure---a critical bottleneck for the entire field~\citep{liang2023holistic,Chiang2023CanLL}.

At the heart of this evaluation challenge lies the \textbf{Oracle Problem} (in this context, the fundamental reliance on an external source of judgment to define ``ground truth'' for agent behavior). This reliance creates a persistent human bottleneck, making scalable and cost-effective evaluation the single most urgent problem in agentic engineering.


To address this, we propose the \textbf{Evaluation-Driven Development Lifecycle (EDLC)} as the formal discipline for building with ArbiterOS. As a specialized application of established MLOps principles (e.g., continuous integration, versioning, and monitoring) to the unique challenges of agentic systems, the EDLC provides a systematic framework for managing the inherent costs and complexities of the ``Oracle Problem.'' It transforms evaluation from a fragile art into a rigorous engineering discipline, centered on a new primary artifact: the \textbf{``Agent Constitution.''} This complete, version-controlled collection of assets is what enables systematic evaluation, and is comprised of three core components:

\begin{itemize}
    \item \textbf{The Execution Graph:} The agent's workflow, defined as a formal graph of ACF Instruction Bindings.
    \item \textbf{Execution Environment Policies:} The declarative, human-readable governance rules that constrain the agent's behavior at runtime.
    \item \textbf{Associated Implementations:} The version-controlled code, prompts, tool definitions, and validators that execute the instructions.
\end{itemize}

This explicit deconstruction is key to enabling systematic evaluation, parallel development (Section~\ref{subsec:Scalability}), and standardized packaging (Section~\ref{subsec:compliance}). Before detailing the lifecycle that operates on this artifact, we first address the cornerstone of its evaluation: the Golden Dataset.

\subsection{The Golden Dataset: A Framework for Amortizing the Cost of the Human Bottleneck}
\label{subsec:golden_dataset}


The EDLC provides the disciplined \textit{process} for evaluation, but the quality of that evaluation is ultimately contingent on the quality of the \textbf{``Golden Dataset.''} The creation and maintenance of this canonical benchmark is where the \textbf{human bottleneck} manifests most acutely. To make this intractable task manageable, we frame the Golden Dataset not as a static artifact, but as a \textbf{``Living Benchmark''} that co-evolves with the agent. The following three-phase process is a strategic discipline designed to \textbf{amortize the non-negotiable cost of human expertise} over the agent's lifecycle, transforming it from a prohibitive upfront burden into a manageable, ongoing investment.

\textbf{Phase 1: Seeding with Domain Expertise (The Upfront Investment).}
The initial dataset is bootstrapped via a direct, upfront investment from the agent's Reliability Budget. For a mission-critical agent, this requires costly time from domain specialists to define critical failure modes and construct detailed golden rubrics. This expert-driven phase is the most expensive but is non-negotiable for establishing a baseline of quality and safety. To scale this initial seed, expert-defined criteria can be used with heuristic techniques like ``LLM-as-a-Judge''~\citep{zheng2023judging} to generate a broader set of test cases, acknowledging the potential for noise that will be refined in subsequent phases.

\textbf{Phase 2: Augmenting from Production Feedback (Amortizing the Cost).}
A core tenet of the EDLC is that no failure is wasted. This principle leverages real-world events to reduce the reliance on manual test case creation, thereby amortizing the cost of evaluation over the agent's lifecycle. When a production failure occurs, the ``Flight Data Recorder'' traces from deployed agents enable systematic, reproducible root-cause analysis. Crucially, each significant production error is then transformed into a permanent regression test and added to the Golden Dataset. This trace-driven workflow converts costly operational firefighting into a disciplined process of incremental, low-cost dataset improvement.

\textbf{Phase 3: Scaling with Adversarial Synthesis (Externalizing the Burden).}
For agents with the highest cost of failure, it is insufficient to rely solely on human curators. To scale beyond the human bottleneck, the Living Benchmark is proactively hardened using automated red-teaming. Dedicated ``attacker'' AI agents are tasked with generating risky or malicious inputs designed to elicit failure~\citep{Perez2022RedTL}. This proactive probing extends the Golden Dataset into uncharted regions of the failure landscape, systematically \textbf{externalizing the burden of failure discovery from human experts to autonomous diagnostic agents}. This phase represents the long-term path to making reliability engineering truly scalable.

\begin{figure}[htbp]
    \centering
    \includegraphics[width=1.0\linewidth]{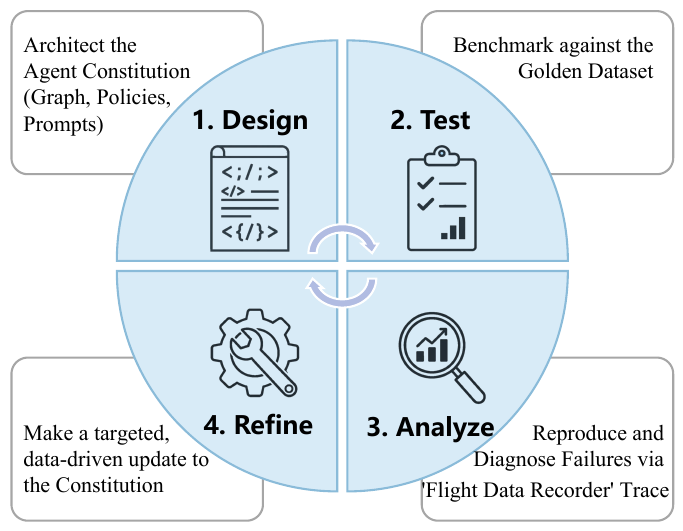}
    \caption{The four phases of the Evaluation-Driven Development Lifecycle (EDLC). This continuous cycle treats the ``Agent Constitution" as the primary artifact, driving measurable improvements in reliability through systematic benchmarking and data-driven refinement.}
    \label{fig:EDLC}
\end{figure}

\subsection{The EDLC Cycle}

With this systematic process for managing the benchmark, we now define the core operational loop of the EDLC. Development is not a linear project but a continuous, four-phase cycle that drives measurable improvements in reliability.

\begin{itemize}
    \item \textbf{Phase 1: Design (Architect the Constitution).} Developers begin the cycle by implementing or modifying the agent's full Constitution—its execution graph, policies, and associated implementations. This is the architectural phase where the agent's formal logic and governance rules are defined.
    
    \item \textbf{Phase 2: Test (Benchmark Against the Golden Dataset).} The complete Agent Constitution is executed against the ``Golden Dataset''—the living benchmark curated through the process described in Section~\ref{subsec:golden_dataset}. This phase generates a comprehensive set of execution traces and performance metrics for every test case, providing the empirical data for the Analyze phase.
    
    \item \textbf{Phase 3: Analyze (Reproduce, Diagnose, and Harden).} This phase focuses on both reactive and proactive analysis. For failures, developers use the ``Flight Data Recorder'' trace to ``time-travel'' debug the exact sequence of state changes, making root-cause analysis a deterministic science rather than guesswork. This same architecture also enables proactive resilience testing, using techniques like systematic fault injection to validate the robustness of fallback plans.

    \item \textbf{Phase 4: Refine (Update the Constitution).} Based on the analysis, the developer makes a targeted, data-driven update to the relevant part of the agent's Constitution. This could involve refining a prompt, hardening a policy, or swapping in a more efficient \texttt{COMPRESS} implementation. Each refinement is a direct, measurable response to an observed or simulated failure, after which the cycle immediately begins again.
\end{itemize}

This continuous, four-phase cycle is the \textbf{engine of agentic engineering}, transforming the abstract goal of ``reliability'' into a concrete, iterative, and empirical process.

Crucially, the interplay between the EDLC and the architecture's HAL provides the only viable engineering response to the `non-stationary hardware' problem. When a new foundation model is released, the HAL allows developers to create new Instruction Bindings (`drivers') while the core Execution Graph remains intact. The EDLC then provides the disciplined process for validating these new drivers against the Golden Dataset, allowing developers to systematically identify and fix behavioral regressions. This transforms the adoption of newer models from a chaotic rewrite into a focused, manageable, and data-driven engineering task, significantly lowering the associated cost, risk, and effort.

%% file: content/Sec7.tex
\section{Situating ArbiterOS: A Governance Framework for the Agentic Ecosystem}
\label{sec:situatingArbiterOS}

\subsection{Platform vs. Ecosystem and the Role of a Unifying Paradigm}
\label{sec:platform_ecosystem}

A simple `layered stack' metaphor is insufficient to capture the complexity of the modern agentic landscape. A more precise taxonomy is to view this ecosystem through four distinct \textbf{Dimensions of Concern}. This model positions ArbiterOS not as another layer in a stack, but as the \emph{Unifying Governance Framework} that brings architectural coherence and principled discipline across all other dimensions.

The agentic ecosystem, including the most advanced platforms from industry leaders, can be mapped onto these primary dimensions:

\begin{itemize}
    \item \textbf{The Execution Dimension}, concerning the low-level mechanics of stateful, graph-based workflows. This is addressed by foundational runtimes like LangGraph and is now robustly implemented in both MAF~\cite{microsoft2025framework} and Google's ADK~\cite{google2025enterprise}.
    
    \item \textbf{The Collaboration Dimension}, focusing on multi-agent interactions and orchestration. This dimension is a key focus of the open ecosystem paradigms, significantly advanced by MAF and Google's ADK through their support for standardized interoperability protocols like A2A/MCP. It is also a core feature of OpenAI AgentKit's multi-agent composition capabilities.

    \item \textbf{The Specification Dimension}, providing declarative languages to describe an agent's design. This is primarily addressed by conceptual frameworks like CoALA, which provide a blueprint for agent architecture.

    \item \textbf{The Tooling \& DX Dimension}, which aims to manage the engineering overhead of agent development. This dimension finds its most advanced expression in the integrated platform paradigms, such as OpenAI's AgentKit, which functions as a `Cognitive IDE' for visualizing and debugging probabilistic workflows, and the observability planes within Google's Gemini Enterprise.
\end{itemize}

While these powerful frameworks provide essential capabilities \textit{within} their respective dimensions, they lack the cross-cutting architectural substrate to guarantee reliability \textit{across} them. ArbiterOS does not compete within these dimensions; it addresses a fifth, orthogonal dimension: \textbf{Governance}. It serves as the unifying framework that integrates the others into a coherent, reliable whole by providing the \textit{Architecture} for the Execution Dimension, the \textit{Robust Citizen Agents} for the Collaboration Dimension, and the \textit{Runtime Enforcement} for the Specification Dimension.

\subsubsection{Key Strategic Approaches in the Market}

This multi-dimensional landscape is currently being shaped by three primary strategic approaches, each representing a distinct philosophy on balancing developer experience, flexibility, and enterprise-grade control:

\begin{itemize}
    \item \textbf{The Integrated Platform (Archetype: OpenAI AgentKit):} The vertical platform approach, offering high ease-of-use and a seamless developer experience. The trade-off is potential ecosystem lock-in.

    \item \textbf{The Open Ecosystem (Archetype: Microsoft Agent Framework):} The horizontal ecosystem approach, offering high flexibility and community extensibility. The trade-off is greater inherent complexity.
    
    \item \textbf{The Hybrid Ecosystem (Archetype: Google Gemini Enterprise):} The hybrid approach, which combines a managed, enterprise-grade runtime with a flexible, open-source SDK. The goal is to provide enterprise-level security and governance without complete ecosystem lock-in, though this can introduce greater complexity than a fully integrated platform.
\end{itemize}

\subsubsection{ArbiterOS as a Foundational Architectural Paradigm}

Ultimately, ArbiterOS is neither a platform nor an ecosystem SDK, but rather a formal \textbf{paradigm and specification} for reliability that transcends both approaches. Its contribution is architectural, not implementational.

Positioned as such, ArbiterOS becomes a foundational component for the entire field. An open-source implementation of the paradigm could serve as a core governance engine within Microsoft's open ecosystem, providing the missing layer of intra-agent reliability. Simultaneously, its \textit{principles} can inform the architectural design of the proprietary kernel underlying OpenAI's platform. This positioning elevates ArbiterOS from a specific implementation to a foundational contribution applicable to both open and closed development models, providing the blueprint for the field's maturation from brittle craft into a robust engineering discipline.

By acting as this unifying governance layer, however, ArbiterOS does more than just fill the gaps in the existing ecosystem. Its formal, integrated structure gives rise to a new set of powerful, emergent capabilities. The following sections detail these second-order benefits, showing how the paradigm enables not only technical reliability but also organizational scalability, advanced engineering practices, and a clear path toward a more mature, standardized, and optimizable agentic ecosystem.

\subsection{Enabling Organizational Scalability}
\label{subsec:Scalability}

This unifying governance framework not only integrates disparate technical dimensions but also enables a new level of organizational scalability. The formal separation of concerns provides a natural blueprint for a clear division of labor, allowing specialized teams to work in parallel within a shared, governable structure.


By deconstructing an agent into its three core components, ArbiterOS clarifies ownership and accelerates development. \textbf{Governance and Safety Teams} can own the declarative \textbf{Execution Environment Policies}, defining safety constraints in human-readable YAML files. \textbf{System Architects} can focus on the high-level \textbf{Execution Graph}, architecting the agent's core logic and workflow. Meanwhile, \textbf{ML and Prompt Engineers} can specialize in the \textbf{Associated Implementations}, refining prompts and tools within the safe boundaries established by the architecture. 

This separation clarifies accountability and transforms debugging, as the ``Flight Data Recorder'' trace can often pinpoint a failure to a specific component, allowing the responsible team to address it swiftly. Ultimately, this clear division of labor represents a socio-technical roadmap for organizations to mature their AI practices: transitioning teams from the heuristic craft of `prompt engineering' to the principled discipline of `constitution architecture,' and fostering a culture of reliability by design.

\subsection{Formalizing Patterns and Best Practices}

The ArbiterOS framework provides the architectural foundation to transform informal agentic patterns into robust, managed processes. The influential ReAct pattern~\citep{yao2023react}, for example, is deconstructed from a simple script into a formal process of \texttt{GENERATE} $\rightarrow$ \texttt{TOOL\_CALL} instructions. This process is then wrapped in systemic protections: an OS-level error handler can catch a failed \texttt{TOOL\_CALL} and trigger a \texttt{FALLBACK}, making the application resilient by design. 

This approach also provides the mechanisms to enforce developer-centric best practices, such as those in the 12-Factor Agent methodology~\citep{humanlayer12FactorAgents}. For instance, the principle that ``Tools are just structured outputs" is no longer a convention but a guarantee, architecturally enforced by the ACF's schema-based bindings~\citep{Bob2021An,Adam2017The}.

\subsection{Enabling Advanced Debugging, Simulation, and Fault Injection}
\label{subsec:advanced_debugging}

The formal structure of ArbiterOS fundamentally enhances the testability and debuggability of agentic systems. 

The combination of a serializable \texttt{Managed State} and a deterministic \texttt{Symbolic Governor} unlocks a suite of powerful engineering practices. This includes \textbf{Time-Travel Debugging}, where the ``Flight Data Recorder" trace enables perfect, step-by-step reproducibility of any failure, drastically reducing the MTTR for production failures. It transforms operational firefighting into a disciplined process of incremental reliability improvement, directly addressing a major source of the human bottleneck in agent maintenance. It also allows for proactive resilience testing through \textbf{Systematic Fault Injection}. Because the kernel mediates all interactions, developers can simulate API failures or other faults to rigorously test governance policies and \texttt{FALLBACK} plans. Finally, the self-contained ``Agent Constitution" acts as a complete \emph{``Digital Twin"} of the agent, enabling high-fidelity simulation in sandboxed environments before high-stakes deployment.

\subsection{Enabling Compliance by Design}
\label{subsec:compliance}

A mature engineering paradigm must also address enterprise-level challenges of regulatory compliance and ecosystem scalability. ArbiterOS is designed to provide the architectural foundation for both.

First, it enables \textbf{Compliance by Design}, transforming compliance from a costly, post-hoc exercise into an intrinsic, verifiable property of the system. This is achieved through two complementary primitives: the declarative \textbf{Policies} provide human-readable evidence of the governance rules, while the immutable \textbf{``Flight Data Recorder"} trace provides the cryptographic-grade audit trail proving those rules were enforced.

Second, the \textbf{Agent Constitution} serves as a standardized, declarative package format—a form of \textbf{``Containerization" for Agents}. Just as Docker standardized software deployment, this allows agents to be packaged in a portable format. This enables a robust ecosystem built on interoperability, with reusable components and portable, sharable ``compliance modules" (e.g., a pre-built policy for GDPR~\citep{GDPR2016}). This transforms how agents are not just built, but also packaged, deployed, and governed at scale.

\subsection{Enabling Provable Safety}
\label{subsec:neurosymbolic}

The ArbiterOS architecture offers a distinct and pragmatic contribution to the field of neuro-symbolic AI. While much research focuses on fusing neural and symbolic capabilities within monolithic models~\citep{d2020neurosymbolic}, ArbiterOS provides a \textbf{compositional approach}. It posits that a highly effective neuro-symbolic system can be built today by composing existing components---a powerful Probabilistic CPU and a deterministic symbolic runtime---and managing their interaction through a formal OS-level contract (the ACF).

This compositional architecture is not merely a design choice; it is the foundational enabler of provable safety and true explainability. By creating a clean, formal boundary between the probabilistic and deterministic components, it opens a crucial, verifiable ``seam'' for the application of \textbf{formal methods}---a capability fundamentally inaccessible to monolithic agent designs. The deterministic Symbolic Governor, its well-defined Execution Graph, and its declarative policies can be formally modeled and verified \textit{before deployment}. Using techniques like \textbf{model checking}, it becomes possible to mathematically prove critical systemic invariants, transforming safety from an aspiration into a guaranteed property.


It is crucial, however, to clarify the scope of this guarantee. Formally verifying the semantic correctness of the Probabilistic CPU's internal reasoning remains an intractable challenge. The promise of ArbiterOS is not to prove that an agent's reasoning is flawless, but rather to prove that its observable actions will \textbf{flawlessly adhere to its constitutional rules}. This distinction is paramount: we are moving the locus of proof from the agent's unobservable ``mind'' to its governable architectural ``body.''

This architectural guarantee is best illustrated with an example. An organization could formally prove a property for a mission-critical financial agent: ``A \texttt{TOOL\_CALL} instruction that interacts with the payments API can \textit{never} be executed unless it is immediately preceded by a successful \texttt{VERIFY} step that checks for transaction limits.'' This is a powerful, provable constraint on the agent's behavior, which is impossible to guarantee for a system whose safety logic is implicitly encoded in a natural language prompt.


Ultimately, this neuro-symbolic separation elevates agent safety from a matter of careful prompt design to a provable architectural feature. This same principle provides a clear solution to the process-level XAI problem: the ``Flight Data Recorder'' trace provides an auditable \textit{`what'} (the symbolic steps), while leaving the Probabilistic CPU to handle the uninterpretable \textit{`why'} (the neural reasoning). Furthermore, concepts like Constitutional AI are no longer soft suggestions but can be implemented as mandatory, non-negotiable \texttt{CONSTRAIN} instructions, rigorously enforced by the OS. All these benefits---provable safety, process-level explainability, and enforceable governance---derive from the same foundational architectural choice.

\subsection{The Enduring Necessity of a Governance Paradigm}
\label{subsec:enduring_necessity}

Beyond situating ArbiterOS within the current ecosystem of tools, it is crucial to address the fundamental tension that justifies its existence as an architectural requirement, rather than an optional tool. This argument forms the core defense for why the ``Abstraction Tax'' of a formal architecture is a necessary investment for the future of the field.

A central tension in agentic AI development lies in balancing rapid iteration with systematic reliability. Much of today's practice can be described as \textit{artisanal craft}: ad-hoc prompt engineering, complex prompt chains, manual orchestration, and brittle recovery mechanisms. From the vantage point of pure agility, the formalism of ArbiterOS might appear as an unnecessary overhead that slows the empirical process responsible for the field's most rapid advances. For academic research and proof-of-concept prototyping, this perspective has merit: speed matters more than predictability.

However, when the objective shifts from demonstrating a capability to deploying a product that must operate reliably in an adversarial world, the artisanal approach fails catastrophically. What begins as nimbleness often degenerates into an unmaintainable web of brittle quick fixes, where the cost of debugging outweighs any initial gain in agility.

ArbiterOS proposes to reframe this dynamic not as an unavoidable penalty, but as a deliberate engineering choice. It acknowledges a fundamental truth: for systems where failure carries real cost, the long-term operational burden of unreliability will always dwarf the one-time cost of abstraction. Therefore, ArbiterOS does not stand in opposition to agile development. Rather, it is the architectural scaffolding that enables successful prototypes to mature into trustworthy products, transforming the ``Abstraction Tax'' into a conscious investment in a governable architecture.

\subsection{Paying Down the Abstraction Tax: A Vision for Constitution-Aware Tooling}
\label{subsec:tooling_vision}

Having established the ArbiterOS paradigm as a necessary architectural investment, we now turn to its second-order benefits: the emergence of a new generation of tooling that can systematically manage and reduce the ``Abstraction Tax.'' The same formal structure that introduces this tax is also the prerequisite for building sophisticated, constitution-aware tools that are inaccessible to artisanal, prompt-based designs. These tools address the two components of the tax directly: \textit{Agent Compilers} to automate performance optimization, and a \textit{Cognitive IDE} to minimize engineering overhead.

\subsubsection{Agent Compilers: Automating Performance Optimization}
\label{subsec:compilers}

Having established \textit{why} this investment is necessary for principled engineering, we now address the practical question of \textit{how} ArbiterOS is designed to manage and systematically reduce its cost.

A primary concern for any OS-level paradigm is the \textbf{Performance Overhead} introduced by the governance layer, particularly its impact on latency. The Arbiter Loop, which intercepts every computational step for validation, imposes a necessary tax. However, the same formal structure that creates this overhead is also the prerequisite for a new class of automated, system-level optimizations that are inaccessible to artisanal designs.

ArbiterOS provides a two-pronged strategy for managing this cost. The first is through dynamic, \textbf{runtime optimizations} enabled by the centralized Governor. Because the kernel has a global view of the execution graph and the formal structure of each instruction, it can perform sophisticated optimizations that individual components cannot:
\begin{itemize}
    \item \textbf{Intelligent Caching:} Caching the results of instructions based on their versioned \texttt{Instruction Binding} and formal input schemas to ensure safe, reproducible cache hits.
    \item \textbf{Parallel Execution:} Analyzing the execution graph to identify and run independent instructions, such as multiple \texttt{VERIFY} checks, concurrently.
    \item \textbf{Dynamic Model Routing:} Acting as a strategic router to send low-stakes instructions to cheaper, faster models, while reserving powerful models for high-stakes cognitive steps, thereby optimizing the agent's cost-performance profile.
\end{itemize}

The second, more powerful strategy is through `compile-time' optimization, which leads to the vision of \textbf{`Agent Compilers.'} Unlike LMP compilers that optimize for task-specific metrics, the primary role of an ArbiterOS compiler is to \textbf{optimize the cost of governance}. These tools would treat the Agent Constitution as an \textbf{Intermediate Representation (IR)} to automatically analyze and rewrite execution graphs for both performance and reliability. For example, a compiler could statically analyze the ``Reliability Budget" to automatically select the most cost-effective ``Gradient of Verification" for a given step, fuse a \texttt{GENERATE} and a Level 1 \texttt{VERIFY} instruction into a single model call while preserving the audit trail, or pre-compile declarative policies into the kernel for near-zero-cost runtime execution.

Ultimately, this two-pronged approach transforms performance tuning from a manual art into a systematic, and potentially automated, engineering discipline.

\subsubsection{The ArbiterOS Toolchain: A Cognitive IDE}
\label{subsec:cognitive_ide}

A formal paradigm is only as good as the tools that make it tractable. To address the \textit{Engineering Overhead} component of the Abstraction Tax, the ArbiterOS paradigm envisions a new generation of constitution-aware tooling, consolidated under the concept of a \textbf{``Cognitive IDE.''} This is not merely a set of optional add-ons, but a foundational requirement for making principled agent engineering an everyday practice. Much as a traditional IDE provides tools for managing code, a Cognitive IDE provides the integrated environment for visualizing, debugging, and governing probabilistic workflows. Its key components would include:

\begin{itemize}
    \item \textbf{A Visual Graph Builder:} An interactive interface for architecting and inspecting the agent's Execution Graph. In this builder, nodes would directly represent formal ACF Instruction Bindings, and edges would represent the control flow. This would make architectural design reviews collaborative, and would make complex error-handling paths and policy boundaries visually explicit.

    \item \textbf{A Policy Editor with Live Linting:} A specialized editor for authoring declarative governance policies in YAML. This tool would provide syntax highlighting, context-aware auto-completion for valid instructions and cores, and, most critically, \textbf{live linting}. By continuously checking the Execution Graph against the active policies, the IDE could immediately flag architectural violations in real-time---for example, warning a developer that a newly-drawn edge from a \texttt{GENERATE} to a \texttt{TOOL\_CALL} node violates a ``think then verify'' policy.

    \item \textbf{An Integrated Time-Travel Debugger:} This is perhaps the most powerful tool enabled by the architecture. By leveraging the serializable ``Managed State'', this debugger would visualize the ``Flight Data Recorder'' trace of any agent execution. It would allow a developer to \textbf{step forward and backward through an agent's entire cognitive process}, inspecting the exact state of \texttt{user\_memory} and \texttt{os\_metadata} at every step. This transforms the debugging of opaque, non-deterministic systems from guesswork into a deterministic science, radically reducing the MTTR for production failures.
\end{itemize}

This dedicated toolchain is what ultimately pays down the Abstraction Tax. It lowers the barrier to adopting constitutionally governed agents and accelerates the transition from rapid prototyping to mature, reliable systems. It is the catalyst that transforms ArbiterOS from theory into disciplined practice.

%% file: content/Appendix.tex
\section{Conceptual Foundations of the ArbiterOS Paradigm}
\label{app:sec:conceptFoundations}

\subsection{The Gradient of Verification and the Reliability Budget}
\label{sec:gradverif}

A central challenge introduced in Section~\ref{sec:computer&budget} is the management of the \emph{Reliability Budget}---the finite engineering resources invested to ensure safe and predictable agent behavior. Reliability, in the context of the \emph{Agentic Computer}, cannot be treated as a binary property; it must instead be conceptualized as a continuum, shaped by probabilistic inference, non-deterministic behaviors, and the absence of perfect oracles for correctness. The question is not \emph{whether} uncertainty can be eliminated, but \emph{how} it should be systematically managed.

\subsubsection{Levels of Verification}

To address this, we introduce the \textbf{Gradient of Verification}, a structured framework that transforms implicit, heterogeneous reliability practices into explicit, auditable engineering decisions. The Gradient conceptualizes verification as existing on a spectrum of rigor, ranging from probabilistic checks to near-deterministic guarantees, as illustrated in Table~\ref{app:tab:VerifyGrad}. This pyramid structure reveals how verification methods trade off cost against guarantee strength—as an agent architect invests more of the Reliability Budget (ascending the pyramid), verification shifts from flexible but probabilistic approaches (Level 1) to rigid but deterministic methods (Level 3), yielding progressively stronger and more auditable guarantees of correctness. Critically, this framing embeds verification into the broader economic context of the Reliability Budget, highlighting that reliability is an \emph{engineered investment strategy} rather than an emergent property.

\begin{table*}[ht]
\centering
\caption{Governance Levels in ArbiterOS: Mechanisms, Investments, and Trade-offs.}
\footnotesize
\renewcommand{\arraystretch}{1.1}
\begin{tabular}{p{2cm} p{2cm} p{6.5cm} p{4.5cm}}
\toprule
\multicolumn{1}{c}{\textbf{Level}} & 
\multicolumn{1}{c}{\textbf{Investment}} & 
\multicolumn{1}{c}{\textbf{Mechanism}} & 
\multicolumn{1}{c}{\textbf{Trade-off}} \\
\midrule
\textbf{Level 1: Soft Governance} & Minimal & Uses an LLM, often guided by a rubric-based prompt, to evaluate the output of another LLM. Capable of assessing nuanced, qualitative properties (e.g., writing style, tone, strategic relevance). Confidence scores (\(p \in [0, 1]\)) can quantify uncertainty. Calibration is improved via Ensemble Verification and temperature scaling. & Flexible, but inherits probabilistic biases and instability (e.g., position bias, sycophancy). Confidence scores must be well-calibrated for reliability. Best for qualitative or high-level semantic checks. \\
\addlinespace
\textbf{Level 2: Hard Governance} & Moderate to high & Traditional deterministic functions (e.g., schema validators, unit tests, regex matches, database queries) provide verifiable, binary \texttt{PASS/FAIL} results. Appropriate for formally specifiable properties. & Strong, trustworthy guarantees, but limited to properties that can be formally specified. Cannot easily assess creative quality or high-level semantics. \\
\addlinespace
\textbf{Level 3: Formal Governance} & Maximum & Escalates verification to definitive ground truth via: (1) \emph{External Oracles}: Authoritative sources (e.g., paid APIs, legal databases). (2) \emph{Human-in-the-Loop}: Explicit human approval (\texttt{INTERRUPT}), pausing execution for review. & Strongest guarantees; highest costs in latency, expense, and human attention. Reserved for critical, non-negotiable checkpoints (e.g., financial transactions, regulatory submissions). \\
\bottomrule
\end{tabular}
\label{app:tab:VerifyGrad}
\end{table*}

The \textsc{ArbiterOS} paradigm ensures guarantees at the level of \emph{process}, not absolute outcome correctness. That is, every verification step is architecturally enforced and rendered auditable. Correctness thus becomes a function of explicit budgeted choices regarding the verification level attached to each instruction. This avoids the pitfall of \emph{governance theater}, where low-rigor checks (e.g., Level~1) are inappropriately deployed in high-cost contexts, providing an illusion of reliability without substantive assurance.

\subsubsection{Managing Residual Risk}
\label{sec:managing_risk}

For domains lacking perfect deterministic oracles, residual risk is unavoidable. \textsc{ArbiterOS} provides two systemic patterns for mitigating this risk:

\begin{enumerate}
    \item \textbf{Ensemble Verification.} Multiple diverse LLM-judges are queried in parallel. Reliability is then derived from consensus, e.g., requiring at least $k$ of $n$ evaluators to agree. The consensus ratio serves as a probabilistic confidence estimate for the Arbiter Loop.
    \item \textbf{Confidence-Based Escalation.} Policies can mandate escalation to higher-rigor checks when confidence scores fall below a threshold. For instance, if a Level~1 verifier outputs $p < 0.8$, execution is paused via \texttt{INTERRUPT} for mandatory human review. This transforms human oversight into a targeted escalation mechanism rather than a constant bottleneck.
\end{enumerate}

\subsection{A Taxonomy of Agentic Systems: Application Archetypes}
\label{sec:taxonomyagent}

Building on the concepts of the \emph{Gradient of Verification} and the \emph{Reliability Budget}, we propose a taxonomy of application archetypes for agentic systems. This taxonomy captures commonly observed design patterns, organizes them according to their reliability requirements, and formalizes the role of governance in their engineering. The archetypes provide a practical bridge between theoretical constructs and real-world system design, clarifying why a configurable operating system such as \textsc{ArbiterOS} is essential.

We identify three canonical archetypes: the \textbf{Executor}, the \textbf{Synthesizer}, and the \textbf{Strategist}. Each is characterized along three dimensions:

\begin{enumerate}
    \item \textbf{Primary Objective:} The functional goal of the system (e.g., correctness, coherence, or strategic success);
    \item \textbf{Reliability Budget:} The justified level of investment in verification and governance, proportional to the cost of failure;
    \item \textbf{Primary Verification Levels:} The typical points along the Gradient of Verification employed to manage risk.
\end{enumerate}

\begin{table*}[ht]
\centering
\caption{Summary of Executor, Synthesizer, and Strategist LLM application archetypes.}
\footnotesize
\renewcommand{\arraystretch}{1.1}
\begin{tabular}{p{1.5cm} p{2.5cm} p{3.5cm} p{7.5cm}}
\toprule
\textbf{Archetype} & \textbf{Domains} & \textbf{Reliability Focus} & \textbf{Core Pattern / Verification} \\
\midrule
Executor & Program synthesis, text-to-SQL, form completion, financial transactions & Very high; correctness and resilience prioritized due to high cost of failure & \texttt{GENERATE} $\to$ \texttt{VERIFY} $\to$ \texttt{EXECUTE}, with registered \texttt{FALLBACK} plans; Level 2 (hard governance) checks (schema validation, tests), Level 3 (formal) for critical steps (\texttt{INTERRUPT}/human-in-loop) \\
\addlinespace
Synthesizer & Summarization, drafting, retrieval-augmented QA, legal memos & Moderate; quality (coherence, factuality, style) prioritized, but failures rarely catastrophic & Iterative refinement: \texttt{GENERATE} $\to$ \texttt{EVALUATE\_PROGRESS} $\to$ \texttt{REFLECT/REGENERATE}; Level 1 (soft governance: rubrics, LLM-as-judge), Level 2 for specific requirements (schema/citation) \\
\addlinespace
Strategist & Autonomous research, project planning, complex web navigation & Variable/adaptive; resilience and metacognition prioritized to avoid invalid strategies & Recursive: $(\texttt{DECOMPOSE} \to \texttt{GENERATE} \to \texttt{EVALUATE\_PROGRESS})^{*} \to \texttt{REPLAN}$; Level 1 (heuristic/soft), Level 2 (resource monitoring); advanced search (tree-of-thought, MCTS) \\
\bottomrule
\end{tabular}
\label{app:tab:archetypes}
\end{table*}

Table~\ref{app:tab:archetypes} synthesizes the three archetypes across their objectives, reliability budgets, and verification practices. The diversity of archetypes underscores the inadequacy of a monolithic, hard-coded agent loop. For Executors, reliability mandates expensive deterministic checks; for Synthesizers, iterative heuristic oversight suffices; for Strategists, dynamic budget allocation prioritizes metacognitive routing and resource governance.  

The configurable execution environments of \textsc{ArbiterOS} are designed precisely for this diversity. Developers can assign distinct governance policies, for example, strict \texttt{VERIFY$\to$EXECUTE} workflows for Executors, rubric-based loops for Synthesizers, and adaptive monitoring for Strategists. This configurability is thus the architectural foundation that transforms agent development from fragile craft into systematic engineering discipline.

\section{Implementation Details of a Governance Paradigm}

\subsection{Declarative Governance in Practice: Policy Examples}
\label{app:sec:policies}

A core tenet of \textsc{ArbiterOS} is the shift from imperative coding to declarative governance: safety and reliability constraints (\emph{what}) are separated from agent logic (\emph{how}). This echoes software best practices such as “configuration over code,” as seen in the 12-Factor App and 12-Factor Agent frameworks~\cite{humanlayer12FactorAgents}.

In \textsc{ArbiterOS}, constraints are defined in external, human-readable policy files (e.g., YAML) and loaded by the Policy Engine at runtime. This modular approach allows safety guarantees to be version-controlled and reviewed independently from agent logic, ensuring governance evolves as a first-class artifact.

\begin{tcolorbox}[colback=black!3!white,colframe=black!30!white,breakable,title=Example: Executor Environment Policy]
\footnotesize

\# Safety constraints for the ``Executor" environment.

\# Primary guarantee: enforce ``think then verify". 
\newline

environment: Executor
\newline

rules:

\quad\quad id: enforce\_verify\_before\_action
  
\quad\quad description: ``Cognitive outputs must be verified before external execution."
    
\quad\quad trigger:
    
\quad\quad\quad\quad \# Activate after any instruction from the Cognitive Core
      
\quad\quad\quad\quad instruction\_core: Cognitive  \# e.g., GENERATE
      
\quad\quad constraint:
    
\quad\quad\quad\quad \# It is a violation if the *next* step is from the Execution Core...
      
\quad\quad\quad\quad violates\_if\_followed\_by\_instruction\_core: Execution  \# e.g., TOOL\_CALL
      
\quad\quad\quad\quad \# ... unless an intervening Normative Core step occurs
      
\quad\quad must\_precede\_core: Normative  \# e.g., VERIFY

\end{tcolorbox}

This policy enforces ``think then verify'': after a Cognitive Core step (e.g., \texttt{GENERATE}), direct execution (e.g., \texttt{TOOL\_CALL}) is only allowed if a Normative Core step (e.g., \texttt{VERIFY}) intervenes. Declarative governance removes safety-critical rules from fragile prompt text and encodes them as auditable, enforceable policies. This decoupling mitigates “prompt drift,” ensuring that changes to prompts do not undermine safety guarantees, and elevates governance to a robust engineering discipline.

\subsection{The Implementation Pathway: Static and Dynamic Validation}
\label{sec:validation}

\begin{figure}[htbp]
    \centering
    \includegraphics[width=0.55\linewidth]{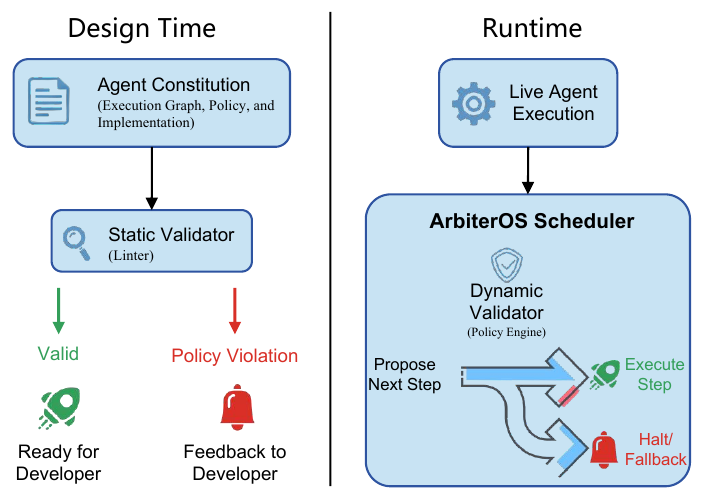}
    \caption{The two-stage validation process in ArbiterOS. At design time, the agent's execution graph is statically analyzed against declarative policies to catch architectural violations before execution. At runtime, the ArbiterOS Scheduler dynamically validates each state transition against the same policies to enforce constraints on conditional, data-driven execution paths.}
    \label{fig:policyValidation}
\end{figure}

A declarative policy is only meaningful if it is rigorously enforced. \textsc{ArbiterOS} achieves this by elevating the Scheduler beyond a passive task runner into an \emph{active validator} of architectural rules. This enforcement occurs through a two-stage validation pipeline, visualized in Figure~\ref{fig:policyValidation}, that exploits the formal, machine-readable structure of the \emph{Agent Constitution} (i.e., the execution graph and its associated ACF Instruction Bindings). The approach mirrors mature practices in software engineering, where a combination of static analysis (\emph{linting}) and runtime checks ensures code quality and security.

\paragraph{Static Validation (Pre-Execution “Linting”).}  
Before execution, an agent’s architecture—represented as a graph of nodes, each a formal ACF Instruction Binding—undergoes static analysis. \textsc{ArbiterOS} offers a validation tool that acts as a linter for agent constitutions. The validator checks the execution graph against the declarative policy of the chosen Execution Environment, systematically tracing all feasible, non-conditional paths to flag structural violations. This process is analogous to a compiler or code linter that catches type errors or security vulnerabilities before execution, providing developers with fast, low-cost feedback through early static enforcement.

\paragraph{Dynamic Validation (Runtime Enforcement).}  
Some violations only emerge at runtime, as conditional and data-driven paths cannot be fully determined statically. For example, the \textsc{Arbiter} Loop may route execution to a \texttt{FALLBACK} step after a \texttt{VERIFY} fails—behavior only revealed during execution. Therefore, validation also occurs dynamically. Before each step, the Scheduler intercepts the proposed state transition, supplying the Policy Engine with contextual metadata from the Managed State (such as the instruction core of the previous and next steps). The Policy Engine then validates this transition against the active policy.

\section{The Agent Constitution Framework (ACF): Full Specification}
\label{sec:acfspecification}

The \textsc{ArbiterOS} paradigm requires not only architectural principles but also a precise formalism
to connect abstract governance concepts with executable system behavior. This is provided by the
\emph{Agent Constitution Framework (ACF)}, a governance-oriented instruction set architecture (ISA).
As introduced conceptually in Section~\ref{sec:acf}, the ACF supplies the formal vocabulary that enables
structured enforcement of safety and reliability guarantees across diverse agent archetypes.

In this section, we provide the technical specification needed for implementation. Section~\ref{app:binding} introduces the \emph{Instruction Binding}
mechanism that operationalizes each ACF instruction as a serializable contract, ensuring inputs, outputs,
and implementations are auditable and enforceable at runtime. Section~\ref{app:instset} then presents the
complete ACF instruction set, organized across the five governance domains (\emph{Cognitive,
Memory, Normative, Execution, and Metacognitive}).

\subsection{The Binding Mechanism: From Instruction to Execution}
\label{app:binding}

The practical force of the ACF comes from its ability to transform abstract design
into enforceable execution. This transformation is achieved through the
\emph{Instruction Binding}, a formal contract that specifies how each ACF instruction is instantiated,
validated, and executed. Conceptually, the binding plays a role analogous to a ``control word" in classical
computer architecture: it provides the canonical interface by which the \textsc{ArbiterOS} kernel
interprets, validates, and governs execution.

\subsubsection{Core Guarantees} 
An Instruction Binding provides three indispensable guarantees:

\begin{enumerate}
    \item \textbf{Formal Typing:} Each instruction is classified according to its ACF domain and type
    (e.g., \texttt{GENERATE}, \texttt{VERIFY}), enabling the policy engine to differentiate between
    probabilistic and deterministic operations.

    \item \textbf{Schema Conformance:} Inputs and outputs are validated against structured, typed
    schemas. This enforces determinism at system boundaries and prevents malformed or adversarial
    outputs from flowing unchecked into external actions.

    \item \textbf{Auditable Execution Reference:} Every Instruction Binding contains a reference to
    a concrete implementation (e.g., a prompt template, a Python validator, or a deterministic tool
    function). This ensures each step is externally traceable and reproducible.
\end{enumerate}

By enforcing strict type validation and schema guarantees, Instruction Bindings act as a
\emph{sanitizing firewall} between the inherently probabilistic operations of the Neural Core
and the deterministic governance of the Symbolic Governor.
This architectural property guarantees the core safety principle of \textsc{ArbiterOS}:
\emph{language models produce structured data, not raw executable commands}.
In turn, the scheduler and Arbiter Loop can treat every step of an agent workflow not as
an opaque chain of thought, but as an auditable, policy-enforceable process.

\subsubsection{Illustrative Example}
The following pseudocode illustrates a binding for a document summarization step.
It encodes the instruction ID, its ACF type, input/output schemas, and the implementation reference. 

\begin{tcolorbox}[colback=black!3!white,colframe=black!30!white,breakable,title=Pseudocode of Instruction Binding]
\footnotesize

INSTRUCTION\_BINDING \{

\quad\quad id: ``summarize\_document'',                  \# unique identifier
    
\quad\quad type: GENERATE,                            \# ACF instruction type (Cognitive Core)
    \newline
    
\quad\quad implementation\_ref: ``prompt/summarize.j2'', \# underlying executor (e.g., template)
    \newline

\quad\quad input\_schema: \{
    
\quad\quad \quad\quad document: STRING                       \# required input
        
\quad\quad \},
    \newline

\quad\quad output\_schema: \{
    
\quad\quad \quad\quad summary: STRING                        \# structured output
        
\quad\quad \}

\}

\end{tcolorbox}

As pseudocode shows, the Instruction Binding abstracts away the low-level details of prompt
content or function calls, and instead foregrounds the \emph{type}, \emph{schemas}, and \emph{implementation
contract}. This allows the \textsc{ArbiterOS} scheduler to reason about the execution graph at the governance level,
ensuring that every transition is validated, logged, and auditable.

\subsubsection{Implications}
Instruction Bindings are therefore not merely implementation details, but the architectural
linchpin that enables \textsc{ArbiterOS} to elevate fragile heuristics into reliable engineering guarantees.
They provide the necessary abstraction for progressive governance: developers incrementally introduce bindings
and associated policies as failure modes are identified, with each binding producing deterministic,
auditable traces that form the basis of evaluation in the EDLC discipline introduced in Section~\ref{sec:discipline}.

\subsection{The Complete ACF Instruction Set}
\label{app:instset}

The \emph{Agent Constitution Framework (ACF)} provides the governance-oriented Instruction Set
Architecture (ISA) through which \textsc{ArbiterOS} exerts control over inherently probabilistic
agentic computation. Whereas the Binding mechanism (Section~\ref{app:binding}) specifies the
contract linking an instruction to its concrete implementation, the ACF itself defines the
\emph{semantic categories of instructions}---the discrete vocabulary required for enforceable,
auditable process governance.

\subsubsection{Rationale}

Unlike traditional orchestrators that treat all computational units
as opaque tasks, \textsc{ArbiterOS} requires semantic granularity. A generic ``LLM call" is
insufficient: a system cannot meaningfully apply distinct governance rules without knowing whether
the operation is content generation, task decomposition, or normative verification. Accordingly,
ACF instructions are defined along two axes: 1) Function: the operational purpose of the instruction in the agent workflow. 2) Governance Property: the trust boundary and architectural obligations imposed by \textsc{ArbiterOS}.

Instructions are grouped into \textbf{\textit{five}} operational domains, reflecting the essential components
of the \emph{Agentic Computer} model: \emph{Cognitive, Memory, Execution, Metacognitive}, and
\emph{Normative}. These domains form the architectural substrate upon which execution
policies are specified and enforced.

\begin{itemize}
    \item \textbf{Cognitive Core (The Mind).}  
    Governs probabilistic reasoning. Its outputs are always treated as
    unverified until subjected to explicit checks. The complete set of Cognitive instructions is listed in Table~\ref{tab:cognitive}.

    \item \textbf{Memory Core (The Context).}  
    Manages the LLM’s limited context window and connections to persistent memory. Detailed instructions are provided in Table~\ref{tab:memory}.

    \item \textbf{Execution Core (The World).}  
    Interfaces with deterministic external systems. These are high‑stakes
    actions requiring strict controls.
    Execution instructions are enumerated in Table~\ref{tab:execution}.

    \item \textbf{Metacognitive Core (The Self).}  
    Enables heuristic self‑assessment and resource tracking, supporting adaptive
    routing in the Arbiter Loop.
    See Table~\ref{tab:metacognitive} for the full instruction set.

    \item \textbf{Normative Core (The Rules).}  
    Enforces human‑defined rules, checks, and fallback strategies. This domain
    anchors \textsc{ArbiterOS}’s claim to systematic reliability.
    The corresponding instructions are detailed in Table~\ref{tab:normative}.
\end{itemize}

\begin{table}[h!]
\centering
\small
\caption{Cognitive Core instructions. These primary instructions invoke the "Probabilistic CPU" for core creative and reasoning functions. Outputs are always probabilistic and require downstream verification.}
\label{tab:cognitive}
\begin{tabular}{p{2cm} p{6cm} p{7.5cm}}
\toprule
\textbf{Instruction} & \textbf{Function} & \textbf{Governance Property} \\
\midrule
\texttt{GENERATE} & Invokes the LLM for text generation, reasoning, or formulating a query. This is the most general-purpose cognitive instruction, producing content including text, hypotheses, and queries. & \textbf{Probabilistic output} that is fundamentally untrusted. The required level and type of verification are determined by the active policy and the criticality of the step. Requires verification before trust or external use. \\
\addlinespace[0.5em]
\texttt{DECOMPOSE} & Breaks a complex task into a sequence of smaller, manageable sub-tasks or creates a formal plan of execution. Transforms complex problems into structured, actionable components. & \textbf{Probabilistic output} where the proposed plan is untrusted. High-reliability policies must validate the plan's structure and feasibility (e.g., via a \texttt{VERIFY} step) before execution to prevent wasted resources and strategic errors. Structural validation required before execution. \\
\addlinespace[0.5em]
\texttt{REFLECT} & Performs self-critique on generated output to identify flaws, biases, and areas for improvement. Often produces a structured critique to guide subsequent \texttt{GENERATE} steps and self-diagnosis of prior outputs. & \textbf{Probabilistic output} where the critique itself is untrusted and may be biased or incomplete. Typically governed by self-correction loops defined in the agent's policy. Output is used to augment user\_memory, not as a trusted signal for the Arbiter Loop. Cannot drive critical transitions directly. \\
\bottomrule
\end{tabular}
\end{table}

Taken together, the ACF instruction set transforms opaque LLM interactions into discrete,
auditable, governance-ready operations. By clearly distinguishing domains and obligations,
\textsc{ArbiterOS} guarantees \emph{process reliability}: even though outcomes remain subject
to probabilistic variability, the enforcement of mandatory checks, compliance filters, and
routing policies ensures that agent executions are transparent, reproducible, and bounded by
a formally auditable constitution.

\begin{table}[h!]
\centering
\small
\caption{Memory Core instructions. These manage the agent's working memory (context window) and its interface with long-term storage. While many are deterministic I/O operations, they can involve probabilistic steps like summarization, which are high-risk.}
\label{tab:memory}
\begin{tabular}{p{2.5cm} p{5cm} p{8cm}}
\toprule
\textbf{Instruction} & \textbf{Function} & \textbf{Governance Property} \\
\midrule
\texttt{LOAD} & Retrieves information from an external knowledge base (e.g., a vector store or document) to ground the agent. & \textbf{Deterministic I/O.} The retrieval process itself is deterministic, but the relevance of the retrieved data is not guaranteed. A subsequent cognitive step is required to utilize the information. \\
\addlinespace[0.5em]
\texttt{STORE} & Writes or updates information in long-term memory, enabling agent learning and persistence. & \textbf{Deterministic I/O.} This is a critical step for agent learning. High-reliability policies may require verification of the data before it is written to prevent memory corruption. \\
\addlinespace[0.5em]
\texttt{COMPRESS} & Reduces the token count of context using methods like summarization or keyword extraction to manage the limited context window. & \textbf{High-Risk Probabilistic Operation.} If implemented with an LLM, this instruction can introduce hallucinations or omit critical data, corrupting the agent's working memory. The output is fundamentally untrusted. High-reliability policies must follow this step with a dedicated \texttt{VERIFY} instruction (e.g., \texttt{VERIFY\_SUMMARY\_FIDELITY} check) to validate the compressed summary against the original information. \\
\addlinespace[0.5em]
\texttt{FILTER} & Selectively prunes the context to keep only the most relevant information for the current task. & \textbf{High-Risk Probabilistic Operation.} If implemented with an LLM, this instruction can incorrectly discard relevant information. The decision is untrusted. Mission-critical policies should include oversight mechanisms, such as requiring a \texttt{VERIFY} step or logging the filtered data for audit, to mitigate the risk of catastrophic information loss. \\
\addlinespace[0.5em]
\texttt{STRUCTURE} & Transforms unstructured text into a structured format (e.g., JSON) according to a predefined schema. & \textbf{Probabilistic Output.} The extracted structure is untrusted and must be followed by a \texttt{VERIFY} step that performs schema validation before it can be used by other components. \\
\addlinespace[0.5em]
\texttt{RENDER} & Transforms a structured data object (e.g., JSON) into coherent natural language for presentation to a user. & \textbf{Probabilistic Output.} The generated text is untrusted and may misrepresent the underlying data. It may require a \texttt{CONSTRAIN} step for style and tone before being shown to a user. \\
\bottomrule
\end{tabular}
\end{table}

\begin{table}[h!]
\centering
\small
\caption{Execution Core instructions. All connect the agent to the external environment and must be preceded by suitable verification.}
\label{tab:execution}
\begin{tabular}{p{2cm} p{6cm} p{7.5cm}}
\toprule
\textbf{Instruction} & \textbf{Function} & \textbf{Governance Property} \\
\midrule
\texttt{TOOL\_CALL} & Executes a predefined, external, deterministic function (e.g., API calls, database queries, code interpreters). Provides structured interaction with vetted external services. & \textbf{Deterministic Action.} Represents direct interaction with the external world. Policies must ensure preceded by appropriate \texttt{VERIFY} checks. Supports sandboxing and requires post-execution verification for critical operations. \\
\addlinespace[0.5em]
\texttt{TOOL\_BUILD} & Writes new code to create novel tools on-the-fly. Enables dynamic capability extension through programmatic generation of custom functions and utilities. & \textbf{High-Risk Probabilistic Action.} Generated code is inherently untrusted and must undergo strict sandboxing and verification (e.g., via \texttt{VERIFY} steps running comprehensive unit tests) before execution permission is granted. \\
\addlinespace[0.5em]
\texttt{DELEGATE} & Passes sub-tasks to specialized agents in multi-agent systems. Facilitates hierarchical task decomposition and leverages domain-specific expertise across agent networks. & \textbf{Deterministic Handoff.} Defines formal, auditable transfer of control and context. While delegation act is deterministic, sub-agent behavior remains probabilistic. OS maintains comprehensive logs of all handoff events for traceability. \\
\addlinespace[0.5em]
\texttt{RESPOND} & Yields final, user-facing output and signals task completion. Serves as the terminal instruction in any execution workflow, ensuring proper task closure. & \textbf{Terminal Action.} Marks definitive task completion. Output must be verified for quality assurance, safety compliance, and factual accuracy before presentation to end-user. Triggers workflow termination protocols. \\
\bottomrule
\end{tabular}
\end{table}

\begin{table}[h!]
\centering
\small
\caption{\textbf{Metacognitive Core instructions.} These instructions enable probabilistic self-assessment and resource monitoring to guide strategic routing decisions within the \textsc{ArbiterOS} Arbiter Loop. They provide heuristic signals for agent introspection rather than deterministic correctness guarantees.}
\label{tab:metacognitive}
\vspace{0.5em}
\begin{tabular}{p{3.2cm} p{5.5cm} p{6.2cm}}
\toprule
\textbf{Instruction} & \textbf{Function} & \textbf{Governance Property} \\
\midrule
\texttt{PREDICT\_SUCCESS} & Estimates the probability of successfully completing the current task or plan, providing anticipatory assessment of feasibility. & \textbf{Probabilistic Self-Assessment.} Produces a heuristic confidence score used for strategic routing (e.g., abandoning low-probability plans). Not a guarantee of correctness. \\
\addlinespace

\texttt{EVALUATE\_PROGRESS} & Performs strategic assessment of the agent's current reasoning path to answer heuristic, goal-oriented questions about viability and productivity. & \textbf{Probabilistic Self-Assessment of Strategy.} Metacognitive check used to detect unproductive paths, escape logical traps, or trigger \texttt{DECOMPOSE}/\texttt{REFLECT} operations. \\
\addlinespace

\texttt{MONITOR\_RESOURCES} & Tracks key performance indicators including token usage, computational cost, and latency against predefined budgets and thresholds. & \textbf{Deterministic Check.} Hard constraint against Reliability Budget limits. Provides trusted PASS/FAIL signals for triggering termination or strategy changes. \\
\bottomrule
\end{tabular}
\end{table}

\begin{table}[h!]
\centering
\small
\caption{Normative Core instructions. These high-privilege instructions form the heart of the Arbiter, providing enforceable correctness and compliance while spending the Reliability Budget.}
\label{tab:normative}
\begin{tabular}{p{2cm} p{4.5cm} p{9cm}}
\toprule
\textbf{Instruction} & \textbf{Function} & \textbf{Governance Property} \\
\midrule
\texttt{VERIFY} & Performs objective correctness checks against verifiable sources of truth (e.g., schemas, unit tests, databases). & \textbf{Deterministic Checkpoint for Correctness.} Primary governance tool providing high-confidence PASS/FAIL signals for critical routing decisions. Implementations range from Level 2 (deterministic code) to Level 1 (LLM-as-judge), but outputs are treated as trusted routing signals. \\
\addlinespace[0.5em]
\texttt{CONSTRAIN} & Applies normative compliance rules ('constitution') to outputs, checking for safety, style, or ethical violations. & \textbf{Architectural Enforcement of Policy.} Implements Constitutional AI concepts. Can be deterministic filter (Level 2) or LLM-based check (Level 1). The OS guarantees execution and enforcement of outcomes. \\
\addlinespace[0.5em]
\texttt{FALLBACK} & Executes predefined recovery strategies when preceding instructions fail (e.g., failed \texttt{TOOL\_CALL}). & \textbf{Deterministic Control Flow.} Provides predefined, trusted recovery paths essential for resilient systems. The Arbiter Loop routes to registered fallback plans based on trusted \texttt{VERIFY} signals. \\
\addlinespace[0.5em]
\texttt{INTERRUPT} & Pauses execution to request human input, preserving agent state for oversight. & \textbf{'System Call' for Human-in-the-Loop (HITL).} Deterministic handoff guaranteeing state preservation and execution halt pending human review. Represents the highest level of governance. \\
\bottomrule
\end{tabular}
\end{table}

\section{Walkthrough Implementation Details}
\label{app:sec:walkthrough_details}

\subsection{Instruction Binding Pseudocode}
The following pseudocode illustrates the formal contracts (\textbf{Instruction Bindings}) for the \texttt{VERIFY} and \texttt{FALLBACK} primitives used in Stage 2 of the walkthrough. These bindings encode the instruction's type, implementation, and I/O schemas, making them governable by the ArbiterOS kernel.

\begin{tcolorbox}[colback=black!3!white,colframe=black!30!white,breakable,title=Pseudocode of Instruction Bindings]
\footnotesize
INSTRUCTION\_BINDING \{

\quad\quad id: ``verify\_api\_response'',            \# unique identifier

\quad\quad type: VERIFY,                             \# ACF instruction type (Normative Core)

\quad\quad implementation\_ref: ``validators.is\_valid\_json\_response'', \# underlying executor

\quad\quad input\_schema: ApiResponseOutput,          \# expected input type

\quad\quad output\_schema: VerificationOutput         \# structured output type

\}
\newline

INSTRUCTION\_BINDING \{

\quad\quad id: ``get\_cached\_sales\_data'',          \# unique identifier

\quad\quad type: FALLBACK,                            \# ACF instruction type (Normative Core)

\quad\quad implementation\_ref: ``cached\_api.get\_cached\_sales\_data'', \# underlying executor

\quad\quad input\_schema: SalesQueryInput,            \# expected input type

\quad\quad output\_schema: ApiResponseOutput          \# structured output type

\}

\end{tcolorbox}

\subsection{Execution Trace: API Outage Scenario}
Table~\ref{tab:api-outage} provides the step-by-step execution log recorded by the \textsc{ArbiterOS} kernel when the agent encounters an external API failure. It demonstrates the deterministic intervention of the Arbiter Loop.

\begin{table}[h!]
\centering
\footnotesize
\caption{Execution trace of the agent under API outage conditions. \textsc{ArbiterOS} ensures graceful degradation by intercepting failed outputs, routing execution through a deterministic fallback pathway, and resuming the primary plan.}
\label{tab:api-outage}
\vspace{0.3em}
\renewcommand{\arraystretch}{1.2}
\begin{tabular}{@{}p{0.6cm}p{4.5cm}p{4.5cm}p{4.8cm}@{}}
\toprule
\textbf{Step} & \multicolumn{1}{c}{\textbf{Instruction Executed}} & \multicolumn{1}{c}{\textbf{Managed State Snippet}} & \multicolumn{1}{c}{\textbf{OS / Arbiter Action}} \\
\midrule
1 & \texttt{TOOL\_CALL} \newline \texttt{(get\_sales\_data)} 
  & api\_response: \newline \texttt{``<html>503...</html>"} 
  & --- \\
\addlinespace[0.5em]
2 & \texttt{VERIFY} \newline \texttt{(verify\_api\_response)} 
  & result: \texttt{``FAIL"}, \newline error\_message: \texttt{``Invalid JSON"}
  & --- \\
\addlinespace[0.5em]
3 & (OS Intervention)
  & ---
  & Arbiter Loop inspects \texttt{FAIL} result in os\_metadata. Routes execution to registered \texttt{FALLBACK} plan. \\
\addlinespace[0.5em]
4 & \texttt{FALLBACK} \newline \texttt{(get\_cached\_sales\_data)} 
  & api\_response: \newline \texttt{\{``sales\_trends": [...]\}} 
  & --- \\
\addlinespace[0.5em]
5 & \texttt{VERIFY} \newline \texttt{(verify\_api\_response)} 
  & result: \texttt{``PASS"}, \newline error\_message: \texttt{null}
  & Arbiter Loop inspects \texttt{PASS} result. Proceeds to next step in the main plan. \\
\bottomrule
\end{tabular}
\end{table}

\subsection{Execution Trace: Strategic Failure Scenario}
Table~\ref{tab:strategic-failure} shows the trace recorded when the agent encounters a reasoning trap. It demonstrates how the \texttt{EVALUATE\_PROGRESS} primitive enables strategic self-correction, enforced by the Arbiter Loop.

\begin{table}[h!]
\centering
\footnotesize
\caption{Execution trace of the agent under strategic failure scenario. \textsc{ArbiterOS} ensures flawed reasoning paths are identified and corrected mid-process, preventing wasted computation and incorrect final conclusions.}
\label{tab:strategic-failure}
\vspace{0.3em}
\renewcommand{\arraystretch}{1.2}
\begin{tabular}{@{}p{0.6cm}p{4.5cm}p{6cm}p{4.3cm}@{}}
\toprule
\textbf{Step} & \multicolumn{1}{c}{\textbf{Instruction Executed}} & \multicolumn{1}{c}{\textbf{Managed State Snippet}} & \multicolumn{1}{c}{\textbf{OS / Arbiter Action}} \\
\midrule
4a & \texttt{GENERATE} \newline \texttt{(summarize\_competitor)} 
   & finding: \texttt{``MegaCorp is} \texttt{a power tool} \newline \texttt{company..."} 
   & --- \\
\addlinespace[0.5em]
4b & \texttt{EVALUATE\_PROGRESS} \newline \texttt{(check\_relevance)} 
   & is\_productive: \texttt{false}, \newline reasoning: \texttt{``Finding is} \texttt{irrelevant to the smart} \texttt{garden market."}
   & --- \\
\addlinespace[0.5em]
4c & (OS Intervention)
   & ---
   & Arbiter Loop inspects false result. Routes execution to the \texttt{REPLAN} node. \\
\addlinespace[0.5em]
4d & \texttt{REPLAN} \newline \texttt{(GENERATE)} 
   & plan: \texttt{``My last search} \texttt{was too broad. I will} \texttt{search for 'MegaCorp} \texttt{smart garden} \texttt{products'..."} 
   & Agent corrects its own strategy. \\
\bottomrule
\end{tabular}
\end{table}

\section{The EDLC in Practice: LLMOps for Agents}
\label{app:edlc}

This section extends the presentation of the Evaluation-Driven Development Lifecycle (EDLC) introduced in Section~\ref{sec:discipline}. While the main body of the paper established the EDLC as a conceptual discipline for governing agentic systems through \textsc{ArbiterOS}, here we examine its operationalization within modern LLMOps infrastructures. Our goal is to show how Continuous Evaluation can be integrated into development workflows, how costs can be systematically managed, and how the \emph{Golden Dataset} can be rigorously curated and maintained.  

\subsection{LLMOps as the Operational Backbone}

Traditional Continuous Integration (CI) pipelines focus on validating software source code using deterministic unit and integration tests. In contrast, the EDLC redefines this practice for the agentic domain as \emph{Continuous Evaluation (CE)}. The core artifact under test is not merely code but the complete \emph{Agent Constitution}---the formal specification of execution graphs, policy files, prompts, and tool bindings.

\subsubsection{Continuous Evaluation Pipeline}
A concrete realization of this process, shown in Figure~\ref{fig:CICDWorkflow}, using CI/CD infrastructure (e.g., GitHub Actions or GitLab CI) can be described as follows:

\begin{figure}[ht]
    \centering
    \includegraphics[width=0.7\linewidth]{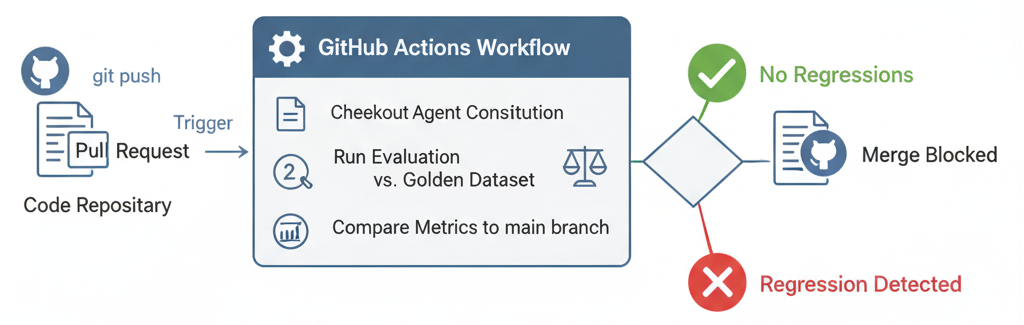}
    \caption{A conceptual Continuous Evaluation pipeline using GitHub Actions. A commit to a pull request automatically triggers a workflow that runs the updated Agent Constitution against the Golden Dataset. If a critical regression in performance is detected, the check fails, automatically blocking the pull request from being merged and preventing behavioral drift.}
    \label{fig:CICDWorkflow}
\end{figure}

\begin{enumerate}
    \item \textbf{Trigger}: Any commit that modifies a constitution artifact (e.g., prompt templates, YAML policies, or tool wrappers) triggers the pipeline.  
    \item \textbf{Checkout}: The full constitution at this commit is retrieved.  
    \item \textbf{Execution}: The agent is run against the version-controlled \emph{Golden Dataset}, producing flight-record traces and metric reports.  
    \item \textbf{Analysis}: Results are compared against the baseline established on the main branch.  
    \item \textbf{Enforcement}: If regressions on reliability metrics (e.g., factuality, task completion, safety adherence) exceed thresholds, the merge is automatically blocked.  
\end{enumerate}

By embedding evaluation as a first-class automation target, \textsc{ArbiterOS} ensures that reliability is not left to artisanal practice but becomes an enforceable property of the software lifecycle.

\subsubsection{Managing the Cost of Continuous Evaluation}

A key challenge with CE is resource overhead: running a full evaluation suite on every commit is computationally and financially intensive. The EDLC addresses this by introducing an explicit \emph{Evaluation Budget}, a sub-allocation of the system’s Reliability Budget (Section~\ref{sec:gradverif}). The following strategies reduce costs while maintaining rigor:

\begin{itemize}
    \item \textbf{Targeted Subset Selection:} Dependency analysis of execution graphs identifies only the test cases relevant to changed components.
    \item \textbf{Deterministic Output Caching:} Expensive, non-probabilistic operations (e.g., embedding calculations, database lookups) are cached across evaluations to avoid redundant computation.
    \item \textbf{Adaptive Evaluation Intensity:} Critical changes (e.g., governance policy updates) trigger full-suite evaluation, while low-risk changes (e.g., stylistic prompt edits) prompt reduced evaluations.
\end{itemize}

This approach transforms evaluation latency and cost from uncontrolled overhead into explicit, budgeted trade-offs aligned with the cost of failure.

\subsection{Budgeting for Trust: The Golden Dataset}

The \emph{Golden Dataset} is the canonical benchmark against which agents are systematically tested. Its form adapts to the archetype in question (Section~\ref{sec:taxonomyagent}):

\begin{itemize}
    \item \textbf{Executor Agents (High Cost of Failure):} Evaluation targets a \textit{Golden Output} using traditional datasets of input-output pairs with deterministic, verifiable outcomes (e.g., mapping a natural language query to the exact SQL query). A large Evaluation Budget supports a comprehensive test suite covering edge cases to ensure correctness.
    \item \textbf{Synthesizer Agents (Moderate Cost of Failure):} Evaluation targets a \textit{Golden Rubric}, with datasets comprising inputs and checklists of qualitative criteria (e.g., “cites three sources,” “tone is neutral,” “avoids speculation”). A moderate Evaluation Budget suffices to define rubrics and curate representative test cases.
    \item \textbf{Strategist Agents (Variable Cost of Failure):} Evaluation focuses on \textit{Golden Guardrails}, assessing the process rather than the final output. The agent is evaluated on goal achievement while adhering to operational constraints (e.g., “booked a trip under budget,” “avoided blacklisted websites,” “did not exceed token limit”).
\end{itemize}

\paragraph{Task-Specific Metrics.}  
Each evaluation modality employs concrete, measurable metrics tailored to the task domain. For example, in retrieval-augmented generation (RAG), a Golden Rubric might use standard reliability measures such as \emph{Faithfulness}, \emph{Answer Relevancy}, and \emph{Contextual Recall}~\cite{Shahul2023RAGAsAE}. Strategic guardrails can be evaluated through resource consumption ratios or constraint satisfaction monitors, while Executor outputs are tested via deterministic schema validation or unit tests.

\paragraph{From Infinite Testing to Finite Discipline.}  
By explicitly linking evaluation design to archetype, budget, and metrics, the EDLC transforms the challenge of “testing an AI agent” from an open-ended problem into a finite, manageable engineering discipline. This targeted, auditable, and budget-conscious approach ensures evaluation is neither prohibitively broad nor overly ad hoc. As the cost of failure increases, the Golden Dataset grows in sophistication, forming the foundation for trustworthy, production-grade systems.

\subsection{A Systematic Process for Dataset Curation}

This archetype-driven approach offers a structured methodology for curating and maintaining the \emph{Golden Dataset}. Human expertise remains essential, but its role shifts from ad hoc, artisanal practices to a formalized, budgeted responsibility within the development lifecycle. This transformation also reveals a core scaling bottleneck in agentic engineering: ongoing dependence on expert-driven failure analysis and manual test creation. While the workflow below marks a significant advance over unstructured practices, it highlights human-centric steps that must ultimately be reduced by targeted automation.

\begin{enumerate}
    \item \textbf{Bootstrap with Domain Expertise:}  
    The initial dataset is built according to the Evaluation Budget. For low-risk prototypes, this may involve a few “happy path” scenarios for basic validation; for mission-critical systems, domain experts and evaluation engineers must invest in defining failure modes, detailed rubrics, and explicit safety guardrails. This ensures early coverage of core risks, though its scale remains limited by expert availability and cost.
    
    \item \textbf{Augment from Production Feedback:}  
    In the EDLC, every significant production error becomes a permanent regression test. Agents are instrumented (e.g., via OpenTelemetry~\cite{OpenTelemetry2025}) to record structured “Flight Data Recorder” traces. These traces provide an auditable account of execution states and routing decisions, enabling systematic root-cause analysis instead of speculative debugging. This workflow reduces Mean Time To Resolution (MTTR) and turns operational firefighting into incremental reliability improvement.
    
    \item \textbf{Synthesize with Adversarial Probing:}  
    For agents with high failure costs, relying only on observed errors is insufficient. Red-team methods—using adversarial LLMs to generate risky or malicious inputs (e.g., prompt-injection attempts, rare edge cases)—proactively surface unknown vulnerabilities. This extends the Golden Dataset into new failure domains, scaling beyond what human curators can anticipate. While adversarial synthesis is early-stage automation, it exemplifies the shift from manual failure discovery to autonomous diagnostics—a key frontier for scalable reliability engineering.
\end{enumerate}

Together, these phases—expert bootstrapping, production feedback, and adversarial synthesis—form a disciplined, evolving process for Golden Dataset curation. Human expertise is progressively redirected from manual debugging to high-level oversight of structured workflows. Embedded in \textsc{ArbiterOS}, this ensures dataset curation remains systematic, auditable, and aligned with the system’s Reliability Budget, transforming evaluation from a fragile art into a scalable engineering discipline.

\section{Ecosystem Context and Detailed Comparisons}
\label{app:ecosystem}

The contributions of \textsc{ArbiterOS} should be viewed within the context of a rapidly evolving ecosystem of agentic frameworks, orchestration mechanisms, and coordination patterns. While existing tools often focus on execution, collaboration, or evaluation, they typically lack the governance substrate necessary for reliable deployment. As discussed in Section~\ref{sec:situatingArbiterOS}, \textsc{ArbiterOS} distinguishes itself as a unifying operating system, providing architectural guarantees and enforcement mechanisms that transform fragile prototypes into governable applications.

To demonstrate this, we highlight influential patterns and algorithms widely adopted in both literature and practice. Although often presented as standalone architectural solutions, these patterns can be seen as \emph{applications} that benefit from the secure and robust execution provided by \textsc{ArbiterOS}'s formal abstractions. Thus, \textsc{ArbiterOS} is not a competitor to emergent practices like \emph{ReAct}~\cite{yao2023react} or \emph{Language Agent Tree Search}~\cite{zhou2024languageagenttreesearch}, but rather serves as the governance framework that enables their reliable production-scale deployment.

\subsection{Detailed Analysis of Agentic Patterns as Managed Processes}
\label{app:patterns}

Two representative cases illustrate the value of interpreting agentic patterns as managed processes
within \textsc{ArbiterOS}: the lightweight reasoning--action loop embodied by \emph{ReAct}, and the
more computationally intensive exploration strategy of \emph{Language Agent Tree Search (LATS)}.

\subsubsection{ReAct as a Governed Process}
The \emph{ReAct} pattern, introduced by \citet{yao2023react}, interleaves natural language
reasoning with external tool calls, typically following the cycle:
\[
\texttt{GENERATE} \;\rightarrow\; \texttt{TOOL\_CALL} \;\rightarrow\; \texttt{OBSERVE}.
\]
\begin{enumerate}
    \item \textbf{Thought:} A \texttt{GENERATE} instruction produces the reasoning step and determines the next action. 
    \item \textbf{Act:} A \texttt{TOOL\_CALL} instruction executes the specified action. 
    \item \textbf{Observation:} The result of the \texttt{TOOL\_CALL} is written back to the central Managed State object, 
    which then becomes the input for the next \texttt{GENERATE} step. 
\end{enumerate}

\begin{figure}[ht]
    \centering
    \includegraphics[width=0.66\linewidth]{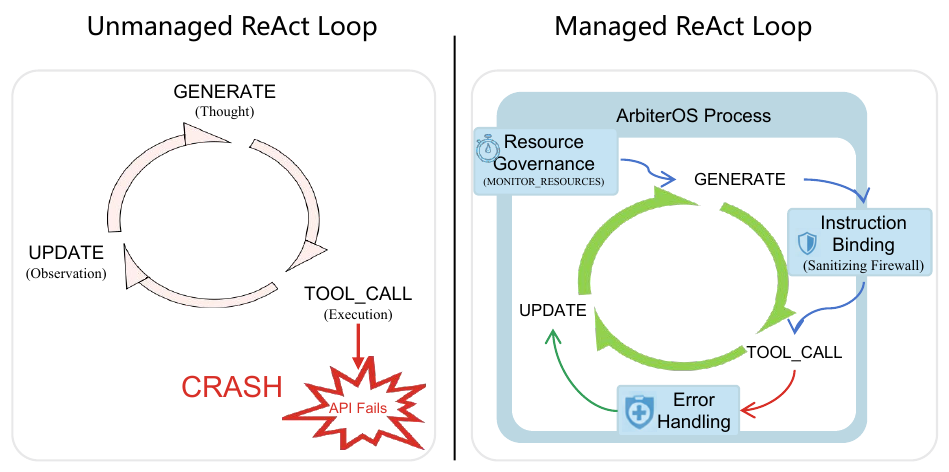}
    \caption{A comparison of an unmanaged ReAct loop versus a ReAct process managed by ArbiterOS. The unmanaged loop is brittle and can crash on a single tool failure. The managed process is wrapped in systemic OS-level protections, providing resilience, security, and resource governance.}
    \label{fig:GovernedReAct}
\end{figure}

While effective as a proof-of-concept, the ungoverned loop is brittle: a single malformed API
response or ill-structured thought may derail execution entirely. When reinterpreted as an
application within \textsc{ArbiterOS}, the pattern benefits from systemic protections as detailed in Figure~\ref{fig:GovernedReAct}:
\begin{itemize}
    \item \emph{Error Containment:} deterministic \texttt{VERIFY} and \texttt{FALLBACK} instructions
    ensure that failed tool calls trigger resilient recovery rather than cascading errors.
    \item \emph{Resource Governance:} \texttt{MONITOR\_RESOURCES} can terminate a runaway ReAct loop and prevents unbounded iteration by
    imposing strict token and time budgets.
    \item \emph{Security Enforcement:} the Instruction Binding acts as a sanitizing firewall, guaranteeing
    that generator outputs are parsed and validated before becoming tool inputs.
\end{itemize}
In this way, \textsc{ArbiterOS} transforms ReAct from an informal recipe into a reproducible,
auditable process suitable for high-stakes deployment.

\subsubsection{Case 2: LATS as a Heavyweight Application}
\emph{Language Agent Tree Search (LATS)}~\citep{zhou2024languageagenttreesearch} exemplifies a heavyweight algorithmic strategy.
It explores a tree of candidate reasoning paths, often using Monte Carlo Tree Search as a guiding heuristic.
While powerful conceptually, implementing LATS without an OS-like framework is extraordinarily challenging:
it requires ad hoc management of search states, error recovery, and bounded exploration.

\textsc{ArbiterOS} makes LATS tractable and governable by providing systemic services:
\begin{itemize}
    \item \emph{State Management:} The OS State Manager tracks the unique Managed State of every node
    in the tree. Each branch is serializable and auditable, and checkpointing ensures fault tolerance
    over long-running searches.
    \item \emph{Node Expansion and Evaluation:} The ACF provides formal verbs for tree search:
    \texttt{DECOMPOSE} generates child plans, and \texttt{EVALUATE\_PROGRESS} scores their promise,
    enabling principled pruning.
    \item \emph{Resource Governance:} \texttt{MONITOR\_RESOURCES} applies non-negotiable global limits,
    pruning exploration paths and terminating runs that exceed token, latency, or budget caps.
\end{itemize}
These systemic services turn LATS from a demanding research prototype into a practical,
auditable application, aligned with explicit Reliability Budgets.

\subsubsection{Case 3: Integrating the Collaboration Dimension}
Multi-agent frameworks such as \emph{AutoGen}~\citep{wu2023autogenenablingnextgenllm} and \emph{CrewAI}~\citep{CrewAIInc2025}
focus on the \emph{Collaboration Dimension}: defining communication protocols, coordination
patterns, and higher-order workflows among specialized agents.
While these approaches excel at orchestrating a ``society of agents", they often assume that each
individual agent is itself reliable and auditable.

Here, \textsc{ArbiterOS} supplies the missing foundation. It ensures that each individual
``citizen" agent is governed, predictable, and composable. For example, an agent built with AutoGen to serve as a 'financial analyst' could be run as a
managed process under an ArbiterOS supervisor. By running subordinate agents under
\textsc{ArbiterOS} supervision, collaboration inherits systemic guarantees:
\begin{itemize}
    \item \emph{Fault Tolerance:} Failed tool calls are caught via \texttt{VERIFY} and repaired with
    \texttt{FALLBACK} plans.
    \item \emph{Auditability:} Every decision is logged as a verifiable Flight Data Recorder trace.
    \item \emph{Resource Management:} Global budgets are enforced via \texttt{MONITOR\_RESOURCES},
    preventing overconsumption or runaway workflows.
    \item \emph{Human Oversight:} High-stakes actions can deterministically trigger \texttt{INTERRUPT},
    allowing human review before continuation.
\end{itemize}

Thus, \textsc{ArbiterOS} is neither redundant with nor in conflict with collaborative frameworks.
Instead, it provides governed components—reliable, auditable citizen agents—that allow a
multi-agent society to function safely at scale.

Across these cases, a general principle emerges: agentic patterns such as ReAct, LATS, and
collaborative societies should not be seen as replacement architectures, but as applications
that become robust only when executed within the \textsc{ArbiterOS} governance framework.
By embedding fragile heuristics in a formally governed substrate, \textsc{ArbiterOS} transforms
them into systematic, auditable, and production-ready processes.

\begin{table}[h]
\centering
\small
\caption{Representative frameworks mapped to the three Dimensions of Concern, and the complementary role played by \textsc{ArbiterOS}.}
\label{tab:frameworks-dimensions}
\vspace{0.5em}
\begin{tabular}{@{}p{2cm}p{3.5cm}p{4cm}p{6cm}@{}}
\toprule
\textbf{Dimension} & \textbf{Primary Concern} & \textbf{Representative\newline Frameworks} & \textbf{Role of \textsc{ArbiterOS}} \\
\midrule
\textbf{Execution} \newline (Mechanism) & 
Efficient execution of cyclical, stateful, or non-deterministic logic; managing control flow and state transitions in agent systems & 
LangGraph~\cite{langchain2023langgraph}; LangChain~\cite{LangChain2025}; basic agent runtimes; Temporal workflows~\cite{LearnTemporal2025}; Apache Airflow~\cite{ApacheAirflow2025} for orchestration & 
Provides \textbf{architecture and governance} layer. The Arbiter Loop and Policy Engine enforce reliability properties atop execution substrates. Offers deterministic state management through Managed State primitives and ensures reproducible execution paths. \\
\addlinespace[0.5em]

\textbf{Collaboration} \newline (Application) & 
Orchestrating interactions and workflows among specialized agents; enabling emergent behaviors through multi-agent coordination & 
AutoGen~\cite{wu2023autogenenablingnextgenllm}; CrewAI~\cite{CrewAIInc2025}; ReAct-style patterns~\cite{yao2023react}; MetaGPT~\cite{hong2024metagptmetaprogrammingmultiagent}; ChatDev~\cite{qian2024chatdevcommunicativeagentssoftware}; CAMEL framework~\cite{li2023camelcommunicativeagentsmind} & 
Supplies \textbf{robust} ``citizen" agents with guaranteed behavioral contracts. Each participant is managed as a governed process---auditable, resource-bounded, and fault-tolerant. Provides inter-agent communication protocols and ensures compositional correctness across agent boundaries. \\
\addlinespace[0.5em]

\textbf{Verification} \newline (Specification) & 
Formal definition of correctness, safety, and behavioral requirements; runtime monitoring and compliance checking & 
CoALA~\cite{sumers2024cognitivearchitectureslanguageagents}; evaluation harnesses; DSPy~\cite{khattab2023dspycompilingdeclarativelanguage} for optimization; Constitutional AI frameworks; RLHF pipelines; formal verification tools & 
Provides \textbf{runtime enforcement and systematic process}. The Policy Engine enforces constitutional ``laws" at execution time, while the EDLC systematizes evaluation as a disciplined process. Enables formal property specification through ACF contracts and maintains audit trails for compliance verification. \\
\bottomrule
\end{tabular}
\end{table}

\subsection{The Ecosystem as a ``Rosetta Stone"}
\label{app:rosetta}

To complement the conceptual foundations of the \textsc{ArbiterOS} paradigm introduced in Section~\ref{sec:taxonomyagent}, 
we provide here two mapping tables that serve as a practical ``Rosetta Stone" for navigating the agentic ecosystem. 
Table~\ref{tab:frameworks-dimensions} situates representative frameworks within the three canonical Dimensions of Concern 
(\emph{Execution}, \emph{Collaboration}, \emph{Verification}), clarifying their primary contribution and 
how \textsc{ArbiterOS} interacts with them. 
Table~\ref{tab:arbiteros-12factor} explicitly links the architectural primitives of \textsc{ArbiterOS} to the 
developer-centric best practices articulated by the 12-Factor Agent methodology~\cite{humanlayer12FactorAgents}. 

Together, these mappings underscore two core claims:
\begin{itemize}
  \item Most leading frameworks address orthogonal slices of the ecosystem rather than competing head-on.
  \item \textsc{ArbiterOS} is not a rival runtime but a unifying governance paradigm that transforms fragile practices into enforceable architecture.
\end{itemize}

\subsubsection{Mapping Frameworks to Dimensions of Concern}
Table~\ref{tab:frameworks-dimensions} reframes the ecosystem from one of rivalry into one of distributed specialization. 
Frameworks such as LangGraph or AutoGen each excel along one conceptual dimension, but they do 
not address system-wide reliability. \textsc{ArbiterOS} acts as the integrative scaffolding, embedding 
these mechanisms within auditable, governed execution environments.

This analysis highlights that \textsc{ArbiterOS} is not positioned as a competitor to existing frameworks, 
but as the integrative operating system that governs them, ensuring cross-dimensional reliability.

\subsubsection{Mapping ArbiterOS Primitives to 12-Factor Agent Principles}
The 12-Factor Agent methodology codifies empirical best practices but lacks systemic enforcement. 
\textsc{ArbiterOS} provides the architectural primitives that embed these principles into runtime, 
analogous to how Kubernetes embedded the 12-Factor App principles into software infrastructure.
Table~\ref{tab:arbiteros-12factor} presents an explicit mapping between \textsc{ArbiterOS} primitives 
and the corresponding 12-Factor Agent principles.

\begin{table}[h]
\centering
\small
\caption{\textsc{ArbiterOS} primitives explicitly mapped to the 12-Factor Agent principles.}
\label{tab:arbiteros-12factor}
\vspace{0.5em}
\begin{tabular}{p{3.5cm} p{4cm} p{8.0cm}}
\toprule
\textbf{ArbiterOS Primitive} & \textbf{12-Factor Principle} & \textbf{Relationship and Enforcement} \\
\midrule
Managed State \& Arbiter Loop & Factor 12: Make your agent a stateless reducer & The OS's core loop implements a pure function that takes current Managed State and an event to produce new state, operationalizing the reducer abstraction with strong reproducibility guarantees. \\[0.5em]
\addlinespace

ACF Instruction Bindings (input/output schemas) & Factor 4: Tools are just structured outputs & The binding mechanism architecturally enforces schema-conformance by ensuring the LLM's role is to produce structured data that conforms to a schema, which is then used to invoke tools with validated inputs. \\[0.5em]
\addlinespace

ACF Instruction Cores (Cognitive vs.\ Execution) & Factor 10: Small, Focused Agents & The formal separation of concerns in the ACF's instruction set encourages composition of agents from small, single-purpose, and verifiable components with clear boundaries. \\[0.5em]
\addlinespace

Policy Engine \& .yaml files & Factor 3: Config (Separate config from code) & Declarative policies externalize governance rules as configuration stored separately from agent's core logic (prompts, tool code), enabling environment-specific behavior without code contamination. \\[0.5em]
\addlinespace

FALLBACK Instruction \& Error Handler & Factor 9: Compact Errors into Context Window & The OS's error handling provides formal structure to catch failures, trigger deterministic FALLBACK plans, and manage how error information is processed and fed back into the agent's state. \\[0.5em]
\addlinespace

INTERRUPT Instruction & Factor 7: Contact humans with tool calls & The OS primitive provides a formal, enforceable 'system call' that guarantees implementation of human oversight best practices, making escalation an auditable event in the system. \\
\bottomrule
\end{tabular}
\end{table}

This mapping reflects a broader pattern in the maturation of engineering disciplines. In the early stages of web development, practitioners relied on informal best practices, which is an ethos analogous to the present role of the 12-Factor Agent methodology. Over time, these informal guidelines were crystallized into enforceable infrastructures such as Kubernetes, which embedded principles like declarative configuration (via YAML) and managed stateless containers into the very substrate of software execution. The same trajectory now unfolds in the agentic domain: the 12-Factor Agent methodology articulates an empirical best-practices ethos, while \textsc{ArbiterOS} instantiates an architectural enforcement paradigm that renders such principles auditable and systematically guaranteed at runtime. The near-simultaneous emergence of these two layers---informal ethos and formal enforcement---is a hallmark of disciplinary maturation, indicating a decisive shift from ad hoc craft toward reproducible and governable engineering of agentic systems.

\subsection{Detailed Distinction from Traditional Workflow Orchestrators}
\label{app:workflow-distinction}

Although the \textsc{ArbiterOS} scheduler utilizes a graph-based state model, resembling workflow orchestrators like Apache Airflow~\citep{ApacheAirflow2025} or Temporal~\citep{LearnTemporal2025}, this similarity is superficial. Traditional orchestrators are designed for reliable completion of \emph{deterministic} workflows, whereas \textsc{ArbiterOS} governs the inherently \emph{probabilistic} substrate of agentic reasoning within the Execution Dimension (see Section~\ref{sec:taxonomyagent}). The distinction is clear across three axes: computational assumptions, representational semantics, and governance capabilities.

\begin{enumerate}
    \item \textbf{Architecture: Probabilistic vs.\ Deterministic Processing}  
    \begin{itemize}
        \item \emph{Traditional Orchestrators:} Assume determinism and reliability—tasks yield identical outputs for identical inputs. Orchestration focuses on scheduling, dependency resolution, handling infrastructure failures, and retrying exceptions.
        \item \emph{\textsc{ArbiterOS}:} Assumes unreliability is endemic. Generative inference can produce hallucinations or errors even with identical inputs. Kernel primitives like the \emph{Arbiter Loop}, \emph{FALLBACK}, and confidence-based routing explicitly model uncertainty, elevating failure-handling to first-class governance under probabilistic conditions.
    \end{itemize}

    \item \textbf{Semantic Instruction Set vs.\ Opaque Operators}  
    \begin{itemize}
        \item \emph{Traditional Orchestrators:} Treat tasks as black boxes, tracking dependencies and I/O without understanding intent. All tasks are equivalent nodes in a graph, regardless of their function.
        \item \emph{\textsc{ArbiterOS}:} Uses the \emph{Agent Constitution Framework (ACF)}, assigning every instruction a formal semantic type (e.g., \texttt{GENERATE}, \texttt{DECOMPOSE}, \texttt{VERIFY}, \texttt{TOOL\_CALL}). This explicit instruction set makes reasoning intent machine-readable, enabling governance at the semantic level.
    \end{itemize}

    \item \textbf{Semantic Governance vs.\ Process Management}  
    \begin{itemize}
        \item \emph{Traditional Orchestrators:} Guarantee process management—task ordering, resource constraints, retry policies—but cannot constrain computational semantics.
        \item \emph{\textsc{ArbiterOS}:} Encodes instruction semantics, allowing machine-enforceable intent-level policies (e.g., requiring a \texttt{VERIFY} after any probabilistic \texttt{GENERATE} before a state-modifying \texttt{TOOL\_CALL}). Such semantic constraints are essential for agentic safety and reliability and have no analogue in Airflow or Temporal.
    \end{itemize}
\end{enumerate}

While both systems employ graph structures, their purposes diverge. Traditional orchestrators ensure the orderly completion of \emph{known deterministic processes}, with correctness defined externally. \textsc{ArbiterOS}, by contrast, governs \emph{unknown probabilistic reasoning paths} arising in agentic computation, focusing on semantic regulation and robust management of uncertain cognitive execution. Thus, \textsc{ArbiterOS} is not a variant of workflow orchestration, but a novel operating system class for probabilistic agents.

\section{Extended Discussion and Future Work}

This section explores the broader implications of the \textsc{ArbiterOS} paradigm and presents a forward-
looking vision for its evolution. The goal is to situate the paradigm within the critical debates shaping the
future of agentic AI and to outline a roadmap for extending its capabilities to meet the next generation
of challenges.

The section is organized into two parts. Section~\ref{appendix:tensions} addresses fundamental tensions in the design space, while Section~\ref{appendix:future_cores} presents a roadmap for novel instruction cores that address emerging challenges in autonomous systems.

\subsection{Extended Discussion on Key Tensions}
\label{appendix:tensions}

Proposing a new engineering paradigm requires more than detailing its architecture; it demands a critical examination of its place within a landscape of competing philosophies. This section addresses the most significant alternative viewpoints that challenge the necessity of a formal, OS-like abstraction layer. We address these not as strawmen, but as legitimate perspectives that help clarify the specific problem ArbiterOS is designed to solve.

\subsubsection{The Price of Reliability: Agile Craft vs. Engineering Discipline}
\label{appendix:reliability_agility}

A central tension in agentic AI development lies in balancing rapid iteration with systematic reliability. 
Much of today’s practice can be described as \textit{artisanal craft}: ad-hoc prompt engineering, complex prompt chains, 
manual orchestration, and brittle recovery mechanisms. Taken at face value, these practices appear fragile and unsustainable. 
Yet a compelling case can be made that this ``craft" is not a defect but a feature of a field unfolding at extraordinary speed.  

\textbf{The Agility Perspective.}  
From this vantage point, the very dynamism of agentic AI demands unconstrained iteration. Formal frameworks, 
the argument goes, introduce an ``Abstraction Tax"—a conceptual and engineering overhead that slows the empirical 
process responsible for the field’s most rapid advances. Premature formalism risks constraining a domain still in 
its \textit{Cambrian explosion}, where novelty and exploration are paramount. For academic research and proof-of-concept prototyping, 
this perspective has merit: speed matters more than predictability, and rapid iteration is the engine of discovery.  

\textbf{The Reliability Imperative.}  
However, when the objective shifts from demonstrating a capability to deploying a product that must operate reliably in 
an adversarial world, the artisanal approach fails catastrophically. The core challenge stems from the friction between two 
fundamentally different paradigms of computation:
\begin{enumerate}
    \item A probabilistic, non-deterministic reasoning core (e.g., large language models) that produces outputs with inherent uncertainty
    \item Deterministic, high-stakes infrastructure (e.g., APIs, databases, ledgers) that demands guaranteed behavioral properties
\end{enumerate}

Attempts to resolve this mismatch through ad-hoc engineering—manual state management, nested prompt chains, patched recovery paths—
inevitably collapse under scale. As system complexity grows, operational fragility grows faster still. What begins as 
nimbleness often degenerates into an unmaintainable web of brittle quick fixes, where the ongoing cost of debugging outweighs any 
initial gain in agility.  

\textbf{Reframing the Trade-off: Reliability Budgets.}  
\textsc{ArbiterOS} proposes to reframe this dynamic not as an unavoidable penalty, but as a deliberate engineering choice: the 
allocation of a \textit{reliability budget}. For low-stakes applications—a personal creative agent, an exploratory research assistant—
the cost of failure is minimal, and the budget for reliability is correspondingly small. In such settings, 
ad-hoc craft remains acceptable, even optimal. By contrast, for mission-critical domains—financial transactions, healthcare decisions, 
autonomous control systems—the budget is necessarily large. Here, the amortized cost of unreliability far exceeds the upfront 
investment in a governable architecture.  

\textbf{A Successor, Not an Enemy.}  
The key insight is that ArbiterOS does not stand in opposition to agile development. Rather, it is the architectural scaffolding 
that enables successful prototypes to mature into trustworthy products. It transforms the "Abstraction Tax" into a conscious, proportional 
investment. In doing so, it acknowledges a fundamental truth: for systems where failure carries real cost, the long-term operational burden 
of unreliability will always dwarf the one-time cost of abstraction.

\subsubsection{Architectural Guarantees vs. Model Capabilities}
\label{appendix:guarantees_capabilities}

A second, more forward-looking position holds that an external OS-like structure will eventually be rendered obsolete by the sheer capability of future foundation models. This \textit{model-centric hypothesis} suggests that next-generation models (e.g., GPT-5 and beyond), trained on vastly more data and with more sophisticated alignment techniques, will internalize reliability, safety, and complex reasoning. In this view, a model might autonomously verify its outputs, gracefully handle tool failures, and adhere to safety constraints as an intrinsic part of its generative process. If realized, such a future could seemingly eliminate the need for a governance layer like \textsc{ArbiterOS}, as the model itself would act as a self-governing entity.

We contend that this perspective, while appealing, conflates two fundamentally distinct categories of system properties:

\textbf{Model-intrinsic properties} concern the quality of reasoning and generation. A future model may achieve near-perfect accuracy in producing safe SQL queries or verifying logical consistency. These capabilities indeed scale with model capacity, training data, and alignment methods. They are probabilistic guarantees about \textit{cognitive performance}.

\textbf{Architectural guarantees}, by contrast, are deterministic invariants baked into the runtime system. They transcend the behavior of any single component. Even a hypothetically perfect reasoning engine must operate in an imperfect environment—facing failing APIs, network timeouts, corrupted data, or hard resource limits. These events are not reasoning failures; they are systemic runtime conditions that demand architectural management.

Consider a concrete example: a model may generate a correct, secure database query with 99.999\% reliability. Yet a mission-critical system requires architectural certainty that:
\begin{enumerate}
    \item Permission verification (\texttt{VERIFY}) is \textit{always} performed before database modification.
    \item Resource constraints (\texttt{MONITOR\_RESOURCES}) are deterministically enforced before expensive operations.
    \item Audit logs capture \emph{all} state-modifying actions for compliance and traceability.
\end{enumerate}

These properties cannot be reduced to probabilistic assurances from a model. They are enforced through deterministic, repeatable mechanisms external to the reasoning core. Expecting the model to provide them is a category error—akin to demanding that a CPU, however advanced, handle the failure of a network card. The principle of \textit{separation of concerns} dictates that no subsystem can be its own final arbiter.

Viewed in this light, \textsc{ArbiterOS} is not a constraint on future models but an enabler. By providing verifiable process guarantees, it reduces the systemic risk of deploying increasingly powerful reasoning engines in high-stakes domains. Far from diminishing in relevance, its architectural role becomes even more critical as models grow more capable. The barrier to deployment at scale is less about model capability and more about safety, predictability, and compliance. In this sense, \textsc{ArbiterOS} is best understood as the scaffolding that allows powerful models to be used responsibly and ambitiously—its value increases, rather than diminishes, with the strength of the models it governs.

\begin{table}[h!]
\centering
\small
\caption{Proposed instructions for the Adaptive Core: Governing autonomous learning and self-improvement within the ArbiterOS paradigm}
\begin{tabular}{p{3.5cm}p{5cm}p{6.5cm}}
\toprule
\textbf{Instruction} & \textbf{Semantic Function} & \textbf{Governance Properties} \\
\midrule
\texttt{UPDATE\_KNOWLEDGE} & Integrates new information into knowledge base via autonomous curriculum generation, web data retrieval, and distillation processes & \textbf{High-Risk Probabilistic Action:} Requires mandatory source verification and \texttt{CONSTRAIN} consistency check before integration due to untrusted knowledge sources \\
\addlinespace[0.5em]
\texttt{REFINE\_SKILL} & Improves existing capabilities through self-generated code testing, fine-tuning on new data, or techniques like Self-Taught Optimizer (STOP) & \textbf{High-Risk Probabilistic Action:} Mandates strict \texttt{MONITOR\_RESOURCES} check and \texttt{VERIFY} step against golden datasets to prevent performance regression \\
\addlinespace[0.5em]
\texttt{LEARN\_PREFERENCE} & Internalizes feedback from human interaction or environmental rewards via Direct Preference Optimization (DPO) or Reinforcement Learning from Human Feedback (RLHF) & \textbf{Probabilistic Self-Modification:} Core decision-making alterations require comprehensive audit trail; high-stakes changes trigger mandatory \texttt{INTERRUPT} for human review \\
\addlinespace[0.5em]
\texttt{FORMULATE\_EXPERIMENT} & Designs and proposes experiments for active learning loops to discover environmental properties or self-capabilities & \textbf{Probabilistic Output:} Untrusted proposals undergo \texttt{PREDICT\_SUCCESS} cost-benefit analysis; high-cost experiments require human approval via \texttt{INTERRUPT} before \texttt{TOOL\_CALL} execution \\
\bottomrule
\end{tabular}
\label{tab:adaptive_core}
\end{table}

\subsection{The Vision for Future ACF Cores}
\label{appendix:future_cores}

Building upon the foundational cores presented in Section~\ref{sec:acfspecification}, we articulate three forward-looking research directions for extending the Agent Constitution Framework (ACF). These extensions address emerging challenges in autonomous systems, particularly in the domains of continual learning, multi-agent collaboration, and human-agent alignment. Collectively, they aim to transform opaque, probabilistic reasoning processes into governable, auditable operations.

\subsubsection{The Adaptive Core: Governed Self-Improvement}
\label{appendix:adaptive_core}

As agents acquire the ability to autonomously learn and refine their capabilities, alignment during adaptation becomes a central concern. We propose the \textit{Adaptive Core}, a sixth instruction set that formalizes learning and self-modification as governable operations. By introducing explicit instructions, the Adaptive Core defines how knowledge acquisition, skill refinement, preference learning, and experimental design can proceed under policy constraints. Table~\ref{tab:adaptive_core} summarizes these proposed instructions and their associated governance properties.

\subsubsection{The Social Core: Multi-Agent Coordination}
\label{appendix:social_core}

The transition from single-agent control to multi-agent ecosystems introduces new challenges of communication, coordination, and emergent risk. To address these, we propose the \textit{Social Core}, which offers a governable instruction set for inter-agent interaction. Its policies ensure that communication protocols are enforced, negotiations remain bounded, and collective decision-making processes yield auditable outcomes. Table~\ref{tab:social_core} presents the proposed instruction set.

\begin{table}[h!]
\centering
\small
\caption{Proposed instructions for the Social Core: Enabling governable inter-agent collaboration in multi-agent systems}
\begin{tabular}{p{3cm}p{5.5cm}p{6.5cm}}
\toprule
\textbf{Instruction} & \textbf{Semantic Function} & \textbf{Governance Properties} \\
\midrule
\texttt{COMMUNICATE} & Sends a structured message to another agent, following a defined protocol for inter-agent coordination & \textbf{Deterministic Handoff.} The OS can enforce communication protocols and log all inter-agent traffic for auditability, though the response remains probabilistic \\
\addlinespace[0.5em]
\texttt{NEGOTIATE} & Engages in a multi-turn dialogue with another agent to reach a mutually acceptable agreement on a resource or plan & \textbf{High-Risk Probabilistic Interaction.} Policies must govern negotiation strategies and enforce timeouts via \texttt{MONITOR\_RESOURCES} to prevent deadlocks or suboptimal outcomes \\
\addlinespace[0.5em]
\texttt{PROPOSE\_VOTE} & Submits a formal proposal to a group of agents and initiates a consensus-forming protocol & \textbf{Formal Consensus Mechanism.} The OS acts as a trusted third party to tally votes and declare an outcome, transforming ambiguous agreement into a discrete, auditable event \\
\addlinespace[0.5em]
\texttt{FORM\_COALITION} & Dynamically forms a temporary group or "crew" of agents to tackle a specific sub-task, defining roles and shared objectives, as seen in frameworks like CrewAI & \textbf{Probabilistic Self-Organization.} The OS can enforce rules on coalition size, composition, and resource allocation to ensure stability and prevent undesirable emergent behavior \\
\bottomrule
\end{tabular}
\label{tab:social_core}
\end{table}

\subsubsection{The Affective Core: Human-Agent Alignment}
\label{appendix:affective_core}

For agents to operate as effective teammates rather than mere tools, they must be capable of reasoning about human cognitive and emotional states. We therefore propose the \textit{Affective Core}, which formalizes socio-emotional reasoning within a governable framework. This enables agents to interpret implicit user intent, adapt communicative behavior, and manage trust dynamics in alignment with human needs. The proposed instruction set is shown in Table~\ref{tab:affective_core}.

\begin{table}[h!]
\centering
\small
\caption{Proposed instructions for the Affective Core: Enabling governed socio-emotional reasoning for human-agent teaming}
\begin{tabular}{p{3cm}p{5cm}p{6.5cm}}
\toprule
\textbf{Instruction} & \textbf{Semantic Function} & \textbf{Governance Properties} \\
\midrule
\texttt{INFER\_INTENT} & Analyzes user communication to infer underlying goals, preferences, or values that may not be explicitly stated & \textbf{Probabilistic Inference.} The inferred intent is a hypothesis, not a fact. High-stakes actions based on inferred intent must be confirmed with the user via an INTERRUPT \\
\addlinespace[0.5em]
\texttt{MODEL\_USER\_STATE} & Constructs or updates a model of the user's current cognitive or emotional state (e.g., confused, frustrated) based on interaction history & \textbf{Probabilistic Self-Assessment.} The model is a heuristic. Policies can use this model to trigger adaptive behaviors, such as simplifying an explanation if the user is modeled as "confused" \\
\addlinespace[0.5em]
\texttt{ADAPT\_RESPONSE} & Modifies a planned response to align with the user's inferred state or established preferences (e.g., adjusting tone, verbosity, or level of detail) & \textbf{Architectural Enforcement of Policy.} This instruction applies a "social constitution" to the agent's output, ensuring its communication style is appropriate and aligned with the user's needs \\
\addlinespace[0.5em]
\texttt{MANAGE\_TRUST} & Evaluates the history of interactions to assess the level of trust the user has in the agent and proposes actions to build or repair that trust & \textbf{Probabilistic Self-Assessment.} A heuristic used for long-term alignment. Can be used to trigger proactive behaviors, like offering more detailed explanations or admitting uncertainty \\
\bottomrule
\end{tabular}
\label{tab:affective_core}
\end{table}

These proposed cores represent critical research frontiers where the \textsc{ArbiterOS} paradigm provides structured approaches to previously intractable problems. Each core extends the fundamental principle established: transforming opaque, probabilistic processes into governable, auditable operations through semantic formalization.





\section{Architectural Support for an ``Era of Experience"}
\label{app:sutton}

The ArbiterOS paradigm, while presented as a methodology for principled agent engineering using today's technology, also offers a compelling architectural bridge to the future of AI. Specifically, it provides the necessary scaffolding to resolve the tension between the pursuit of experience-based learning systems and the non-negotiable requirement for their safe and reliable operation. This is best framed by the perspective of reinforcement learning pioneer Richard Sutton and his influential critique of knowledge-based approaches~\cite{Sutton2025}.

\subsection{Sutton's Vision: The Primacy of Experience}

Sutton's arguments, rooted in his seminal 2019 essay ``The Bitter Lesson,"~\cite{Sutton2019} are a powerful call to re-center AI research on scalable, general-purpose methods. His position, which challenges the architectural foundations of many contemporary systems, can be summarized as follows:
\begin{itemize}
    \item \textbf{The Bitter Lesson:} The greatest long-term gains in AI have consistently emerged from general methods that leverage computation (namely, search and learning), not from systems into which complex, handcrafted human knowledge has been encoded.
    \item \textbf{The Critique of Mimicry-Based Systems:} Current LLMs are critiqued as sophisticated ``mimicry engines." Trained on a static dataset of human text, they learn to predict a human-like response. Sutton argues this approach is fundamentally limited because it lacks a true goal or a grounding in real-world consequences; it models \textbf{what a human would say, not what will happen}.
    \item \textbf{The Primacy of Experience:} True intelligence must be grounded in experience. He contends that agents must learn continually by interacting with an environment, receiving feedback (rewards), and updating their own world models. This defines the coming ``\textbf{Era of Experience}."
    \item \textbf{The Need for Meta-Methods:} The goal of AI research should not be to build in our discoveries about the world, but to build in the \textit{meta-methods} that enable an agent to discover for itself.
\end{itemize}
Sutton's core challenge is that current architectures cannot learn continually from experience without suffering from catastrophic forgetting or alignment failures. He posits that a new architecture is needed to enable this on-the-fly, experience-driven learning in a stable manner.

\subsection{ArbiterOS as the Governance Layer for Continual Learning}

The ArbiterOS paradigm provides a concrete architectural response to this challenge. While pragmatically designed to govern today's LLM-based agents, its foundational separation of concerns is perfectly aligned with enabling Sutton's ``Era of Experience."

ArbiterOS is, in essence, a framework for building the very ``meta-methods" Sutton calls for. Its neuro-symbolic design separates the agent into two components: the untrusted, probabilistic learner (the ``System 1" \textit{Probabilistic CPU}) and the trusted, deterministic governor (the ``System 2" \textit{Symbolic Governor}). This architectural separation is the key. It creates a safe execution environment where a learning component—be it an LLM or a future reinforcement learning (RL) model—can experiment and learn from experience, while the \textit{Symbolic Governor} ensures the learning process adheres to systemic safety, ethical, and operational constraints defined in its policies.

This vision is most directly supported by the future-facing extensions to the Agent Constitution Framework (ACF), which provide the specific, governable primitives required to implement experience-based learning.

\subsubsection{The Adaptive Core: Primitives for Governed Self-Improvement}

The proposed \textbf{Adaptive Core} is the most explicit architectural hook for Sutton's vision. It formalizes learning and self-modification as governable, auditable operations. Its instructions provide the ``meta-methods" for learning, while the ArbiterOS kernel provides the enforcement.
\begin{description}
    \item[UPDATE\_KNOWLEDGE] Allows an agent to integrate new information. The governor can enforce a policy requiring this instruction to be preceded by a \texttt{VERIFY} step that checks a source's credibility, preventing the agent from learning from un-vetted or malicious data.
    \item[REFINE\_SKILL] Enables the agent to improve capabilities through practice. The governor can use a \texttt{MONITOR\_RESOURCES} check to budget this process and a \texttt{VERIFY} step against the Golden Dataset to prevent performance regressions, ensuring the agent does not ``un-learn" a critical skill.
    \item[LEARN\_PREFERENCE] Formalizes the internalization of feedback from human interaction or environmental rewards—the core of RL. The governor can be configured with a policy that flags high-stakes changes to the agent's core preferences and triggers a mandatory \texttt{INTERRUPT} for human review, providing a critical alignment safety mechanism.
\end{description}
These primitives transform learning from an opaque, uncontrollable process into a set of discrete, auditable, and policy-driven operations. This directly addresses the challenge of building agents that can ``discover like we can," by providing the safe framework in which that discovery can occur.

\subsubsection{The EDLC as a Formalized Learning Loop}

On a macro scale, the Evaluation-Driven Development Lifecycle is itself an implementation of an experience-based learning loop:
\begin{enumerate}
    \item \textbf{Act:} The agent's \textit{Agent Constitution} is executed against the \textit{Golden Dataset}.
    \item \textbf{Observe:} The \textit{Flight Data Recorder} captures a high-fidelity trace of the agent's performance—its ``experience."
    \item \textbf{Learn:} In the Analyze and Refine phases, this experience is used to make targeted, data-driven updates to the agent's constitution, improving its performance.
\end{enumerate}
This cycle, mirroring the core loop of RL, provides the discipline to ensure that learning from experience is systematic, measurable, and drives tangible improvements in reliability.

\subsection{Conclusion: A Bridge to the Era of Experience}

Sutton is correct that true intelligence will likely emerge from systems that learn continually from interaction with the world. However, deploying autonomous, self-modifying agents without a robust framework for governance is an unacceptable risk.

ArbiterOS resolves this tension. It is not an ``LLM framework"; it is a \textbf{governance framework for probabilistic processors}, whatever their nature. It provides the architectural separation, formal instruction set (the ACF), and rigorous development discipline (the EDLC) necessary to build agents that can learn from experience safely.

By providing primitives like the Adaptive Core, ArbiterOS offers a concrete engineering pathway to implement Sutton's vision. It allows us to build agents capable of entering the ``Era of Experience" and learning for themselves—all while remaining auditable, reliable, and aligned with human-defined rules. In this light, ArbiterOS is not a counterargument to Sutton's vision, but one of its most critical enablers.